\documentclass[journal]{IEEEtran}
\IEEEoverridecommandlockouts
\usepackage{cite}
\usepackage{overpic}
\usepackage{amsmath,amssymb,amsfonts}

\usepackage{comment}

\usepackage{graphicx}
\usepackage{textcomp}
\usepackage{xcolor}
\usepackage{float}
\usepackage{amsthm}
\usepackage{graphicx}
\usepackage{epstopdf}
\usepackage{amsmath,bm}
\usepackage{amsfonts}
\usepackage{amssymb}
\usepackage{color}
\usepackage{multirow}
\usepackage{multicol}
\usepackage{soul,xcolor}
\usepackage{algorithm}
\usepackage{algpseudocode}

\newcommand{\mathacr}[1]{\mathsf{#1}}
\theoremstyle{plain}

\newtheorem{lemma}{Lemma}

\newtheorem{corollary}{Corollary}
\newtheorem{remark}{Remark}

\newcommand{\vect}[1]{\mathbf{#1}}

\def\diag{\mathrm{diag}}

\def\kron{\otimes}
\def\tr{\mathrm{tr}}
\def\rank{\mathrm{rank}}
\def\Htran{\mbox{\tiny $\mathrm{H}$}}
\def\Ttran{\mbox{\tiny $\mathrm{T}$}}
\def\CN{\mathcal{N}_{\mathbb{C}}} 
\def\imagunit{\mathsf{j}} 

\def\sinc{\mathrm{sinc}}

\begin{document}

\title{Efficient Channel Estimation With Shorter Pilots in RIS-Aided Communications: Using Array Geometries and Interference Statistics}

\author{\IEEEauthorblockN{\"Ozlem Tu\u{g}fe Demir, \emph{Member, IEEE}, Emil Bj{\"o}rnson, \emph{Fellow, IEEE}, Luca Sanguinetti, \emph{Senior Member, IEEE}\vspace{-0.9cm}}
\thanks{
\"O. T. Demir is with the TOBB University of Economics and Technology, Ankara, T\"urkiye (e-mail: ozlemtugfedemir@etu.edu.tr.) E.~Bj\"ornson is with the School of Electrical Engineering and Computer Science, KTH Royal Institute of Technology, Stockholm, Sweden (e-mail: emilbjo@kth.se). L.~Sanguinetti is with the University of Pisa, Dipartimento di Ingegneria dell'Informazione, 56122 Pisa, Italy (e-mail:luca.sanguinetti@unipi.it). \newline \indent This work was supported by the FFL18-0277 grant from the Swedish Foundation for Strategic Research. The work of Luca Sanguinetti was supported by the Italian Ministry of Education and Research (MUR)
in the framework of the FoReLab project (Departments of Excellence) and the Garden Project (PRIN 2022 program).
\newline \indent A preliminary version of this paper was presented at the IEEE 12th Sensor Array and Multichannel Signal
Processing Workshop (SAM) \cite{demir2022exploiting}.}
}

\maketitle

\begin{abstract}
 Accurate estimation of the cascaded channel from a user equipment (UE) to a base station (BS) via each reconfigurable intelligent surface (RIS) element is critical to realizing the full potential of the RIS's ability to control the overall channel. The number of parameters to be estimated is equal to the number of RIS elements, requiring an equal number of pilots unless an underlying structure can be identified. In this paper, we show how the spatial correlation inherent in the different RIS channels provides this desired structure.  
We first optimize the RIS phase-shift pattern using a much-reduced pilot length (determined by the rank of the spatial correlation matrices) to minimize the mean square error (MSE) in the channel estimation under electromagnetic interference. In addition to considering the linear minimum MSE (LMMSE) channel estimator, we propose a novel channel estimator that requires only knowledge of the array geometry while not requiring any user-specific statistical information. We call this the \emph{reduced-subspace least squares (RS-LS)} estimator and optimize the RIS phase-shift pattern for it. This novel estimator significantly outperforms the conventional LS estimator. For both the LMMSE and RS-LS estimators, the proposed optimized RIS configurations result in significant channel estimation improvements over the benchmarks.
\end{abstract}

\begin{IEEEkeywords}
RIS, channel estimation, reduced-subspace least squares, spatial correlation matrix, pilot design, electromagnetic interference. \vspace{-0.5cm}
\end{IEEEkeywords}

\section{Introduction}

Reconfigurable intelligent surface (RIS)-aided communication is a pivotal area of investigation for next-generation wireless systems \cite{Wu2019,RISchannelEstimation_nested_knownBSRIS2,pei2021ris}. An RIS is a planar array of $N$ reflecting elements (meta-atoms) with sub-wavelength spacing. Each element can be configured by adjusting its impedance to induce a controllable phase-shift of the incident wave prior to reflection. By optimizing the phase-shift pattern across the RIS, the reflected wavefront can be shaped (e.g., as a near-field/far-field beam toward the intended receiver) \cite{Bjornson2022a}. Estimating the channel coefficients associated with each element is a key challenge, as RISs are envisioned to consist of hundreds of elements \cite{RIS_emil_magazine}.

In a basic RIS-aided communication setup with a single-antenna base station (BS), a single-antenna user equipment (UE), and $N$ RIS elements, the channel is described by $N+1$ coefficients. These coefficients represent the controllable channels via the respective elements plus the uncontrolled direct BS-UE channel. Classical channel estimators (e.g., the least squares (LS) estimator) require a pilot length of at least $N+1$ to function properly. The works on passive RIS-aided channel estimation can be broadly classified into two branches \cite[Tab.~2]{xu2023reconfiguring}. The first branch employs ON/OFF techniques \cite{mishra2019channel,you2020channel,alwazani2020intelligent,wang2020channel}, which require a large pilot overhead. Conversely, low pilot overhead is achievable when the RIS channel is sparse \cite{he2019cascaded,xia2020intelligent,de2020parafac,yuan2022channel,wei2021channel}. However, reducing pilot overhead remains a challenge when the channel lacks a sparse structure.

Various methods that exploit sparsity, spatial channel correlation, and/or other specific characteristics of the channel can be employed to improve channel estimation beyond classical estimators \cite{ris_training}. One such technique involves exploiting spatial correlation among the channel coefficients. In \cite{ris_channel_estimation_lmmse, RISchannelEstimation_nested_knownBSRIS2,ris_joint_training_phase}, the authors minimize either the mean square error (MSE) or the effective noise variance of the linear minimum MSE (LMMSE) estimator. However, these methods are only applicable when the pilot length is at least $N+1$. Additionally, sparsity inherent in mmWave and THz channels can be leveraged to design parametric channel estimation methods with low overhead \cite{wu2023parametric,he2021channel,liu2020matrix,zhou2022channel}. For example, in \cite{wu2023parametric}, the authors develop a channel estimation method to obtain sparse multipath components such as angles, distances, and path gains of the near-field THz channel. Similarly, \cite{he2021channel} uses a sequential estimation approach for angle parameters, angle differences, and products of propagation path gains. The authors of \cite{liu2020matrix} propose a message-passing-based channel estimation algorithm by exploiting the slow-varying channel components and hidden channel sparsity. In \cite{zhou2022channel}, the sparsity and correlation of multi-user mmWave channels are exploited to design low-overhead channel estimation. In \cite{haghshenas2023parametric}, a parametric maximum likelihood estimation framework is presented for estimating the line-of-sight channel between UE and RIS.

Recently, it has been shown that reducing the pilot length is possible by equipping the RIS with a radio frequency (RF) chain to empower it with a transmission capability \cite{zhu2023channel}. However, for a fully passive RIS, the challenge of low pilot overhead still exists in a rich scattering environment. None of the above parametric approaches consider a rich scattering environment. To the best of our knowledge, this paper is the first to optimize the RIS phase-shift configuration during channel estimation by exploiting the spatial channel correlation to reduce the MSE for a given pilot length, while simultaneously exploiting the reduced-rank properties caused by the array geometry to enable shorter pilot lengths than $N+1$.

In contrast to earlier studies primarily leveraging array geometry for the development of low-overhead channel estimation schemes, our research delves into a more comprehensive exploration. We not only exploit the array geometry but also address the detrimental impact of electromagnetic interference (EMI) on the RIS during pilot transmission. This EMI, resulting from directional or isotropic interference from other UEs \cite{long2023channel}, has been identified as a significant factor affecting communication performance, especially as the RIS size increases \cite{Torres2021}. EMI can occur accidentally due to various uncontrolled electromagnetic sources, including natural pollution/radiation and human-made devices or intentionally \cite{Torres2021,chandra2022downlink,vega2022physical}. Since the RIS also reflects EMI, its effect should be analyzed rigorously. While the authors of \cite{chandra2022downlink,hassouna2023reconfigurable,long2023mmse} focus on the performance of RIS-based communications under EMI, \cite{khaleel2023electromagnetic} proposes an EMI cancellation method that accounts for slowly changing EMI.

It is important to note that the majority of prior works, except for \cite{long2023channel,long2023mmse}, do not specifically address channel estimation. In \cite{long2023channel}, the method presented in the conference version of this paper (i.e., \cite{demir2022exploiting}) is employed in a two-stage training framework. In the first stage, the BS exclusively estimates EMI, assuming it remains constant during the pilot phase. Although our analysis shares this assumption, it offers a suboptimal solution in scenarios where EMI changes dynamically during pilot transmission. We thoroughly evaluate this performance in our numerical results.

In \cite{long2023mmse}, both LMMSE-based channel estimation and MMSE-based data detection under the influence of EMI are considered. The authors propose alternating optimization algorithms to optimize the RIS phase-shift pattern, ensuring local optimality. In our research, we extend this by providing the global optimal structure of the RIS phase-shift, omitting unit-modulus constraints. This extension yields a lower bound on the achievable mean-square error (MSE).

Our contributions can be outlined as follows. We consider the spatial correlation seen at both the BS and the RIS and provide the corresponding LMMSE channel estimator. Inspired by our recent work without RIS \cite{demir2022channel}, we propose a novel channel estimator that exploits the reduced-rank subspace created by the array geometry to improve the estimation quality without requiring UE-specific spatial correlation information---which is difficult to obtain in cases with a large number of BS antennas and RIS elements. The proposed \emph{reduced-subspace least squares (RS-LS)} estimator outperforms the conventional LS estimator and enables reduced pilot training time. For both the LMMSE and RS-LS estimators, we derive the ideal RIS configuration that minimizes the respective MSE. We then demonstrate by simulation that projecting to the closest unit modulus RIS responses provides significantly better performance than the benchmarks. Furthermore, as a superior alternative, we propose a majorization-minimization (MM) algorithm for designing the RIS phase-shift matrix to minimize the MSE of the RS-LS estimator, offering a more effective approach to configuring the RIS for improved system performance. Unlike the conference version \cite{demir2022exploiting}, which focuses solely on RS-LS without EMI, this work provides a comprehensive analysis of both the LMMSE and RS-LS estimators under EMI through extensive mathematical analysis, demonstrating the robustness and applicability of our methods in practical EMI conditions.

\subsubsection*{Paper Outline} Section~\ref{sec:system} presents the channel and system modeling for RIS-aided communications. The pilot transmission under EMI and the proposed LMMSE- and RS-LS-based channel estimators are described in Section~\ref{sec:pilot}. The optimal structure of the phase-shift configuration with and without EMI is derived in Section~\ref{sec:optimization-LMMSE} for the LMMSE estimator. The same is done with the RS-LS channel estimator later in Section~\ref{sec:optimization-RSLS}, where a suboptimal structure is obtained under EMI. Section~\ref{sec:numerical} presents a number of interesting numerical results, and Section~\ref{sec:conclusions} offers final thoughts to wrap up the work.

{\bf Reproducible research:} All the simulation results can be reproduced using the Matlab code available at:
https://github.com/ozlemtugfedemir/RIS-shorter-pilots

\section{System and Channel Modeling} \label{sec:system}

We consider the uplink communication between a single-antenna UE and a multi-antenna BS with the aid of an RIS. The BS has $M$ antennas and the RIS has $N$ reconfigurable elements. The BS antennas are deployed as a uniform planar array (UPA) with $M_{\rm H}$ elements per row and $M_{\rm V}$ elements per column. Hence, $M=M_{\rm H}M_{\rm V}$. Similarly, the RIS elements form a UPA with $N_{\rm H}$ and $N_{\rm V}$ number of elements per row and per column, respectively, so that $N=N_{\rm H}N_{\rm V}$.  
Each RIS element is passive and introduces a phase-shift to the signals that impinge on it before reflection~\cite{Wu2019,RISchannelEstimation_nested_knownBSRIS2,pei2021ris}.

We consider a time-varying narrowband channel and adopt the conventional block fading model  \cite[Sec. 2.1]{massivemimobook}, where the time resources are divided into independent coherence blocks. The channel responses are constant within each coherence block, and thus are represented by complex scalars. We let $\tau_p$ denote the total number of samples allocated to pilot transmission per coherence block. The channel from the UE to the RIS is denoted by $\vect{h} = [h_1,\ldots,h_N]^{\Ttran}\in \mathbb{C}^N$. In a UPA, there is always spatial correlation between the channel coefficients observed at the different elements \cite{Bjornson2021b,demir2022channel}. Hence, we adopt a spatially correlated Rayleigh fading model, i.e., $\vect{h}\sim\CN(\vect{0}_N,\vect{R}_{\rm h})$, where $\vect{h}$ takes an independent realization in each coherence block. We adopt the general spatial correlation matrix model~\cite{Sayeed2002a}
\begin{equation} \label{eq:spatial-correlation}
\vect{R}_{\rm h} = \beta_{\rm h}  \iint_{-\pi/2}^{\pi/2} f_{\rm h}(\varphi,\theta) \vect{a}(\varphi,\theta) \vect{a}^{\Htran}(\varphi,\theta) d\theta d\varphi 
\end{equation}
where $\beta_{\rm h}\geq0$ is the channel gain, $\vect{a}(\varphi,\theta)\in \mathbb{C}^N$ denotes the array response vector, while $\varphi$ and $\theta$ are the azimuth and elevation angles, respectively.
The normalized spatial scattering function is called $f_{\rm h}(\varphi,\theta)$~\cite{Sayeed2002a}. We assume a coordinate system where the RIS is deployed in the $yz-$plane and the waves only arrive from directions in front of it; that is, $f_{\rm h}(\varphi,\theta)$ is only non-zero for $\varphi, \theta \in[-\frac{\pi}{2},\frac{\pi}{2}]$.

The rank and subspaces of $\vect{R}_{\rm h}$ are determined by the array geometry and element spacing through $\vect{a}(\varphi,\theta)$, but also by the non-isotropic scattering and directivity pattern of the elements through $f_{\rm h}(\varphi,\theta)$ \cite{demir2022channel, Demir2021RIS}. The largest rank\footnote{The rank is computed as the effective rank, containing a fraction $1-10^{-6}$ of the sum of all eigenvalues throughout this paper. \label{footnote1}} is obtained in an \emph{isotropic scattering environment} where the multipath components are uniformly distributed over all directions and the antennas are isotropic. This can be represented by the spatial scattering function\footnote{The cosine of the elevation angle, $\cos(\theta)$ appears in the expression since it is the differential of the solid angle in the spherical coordinate system.}  $f_{\rm h}(\varphi,\theta) = \cos(\theta) / (2\pi)$ and the $(m,l)$th entry of the resulting normalized correlation matrix $\vect{R}_{\rm iso}$  is~\cite{Bjornson2021b}
\begin{equation}\label{R-iso}
    \left[\vect{R}_{\rm iso} \right]_{m,l} =  \sinc \left( 2 \sqrt{\left(d_{\rm H}^{ml}\right)^2+\left(d_{\rm V}^{ml}\right)^2}\right).
\end{equation}
In this expression, $\sinc(x) = \sin(\pi x)/ (\pi x)$ is the sinc function, while $d_{\rm H}^{ml}$ and $d_{\rm V}^{ml}$ denote the horizontal and vertical distances (normalized by the wavelength) between antenna (or RIS element) $m$ and $l$, respectively. 

We let $g_{m,n}\in \mathbb{C}$ denote the channel from RIS element $n$ to BS antenna $m$, for $n=1,\ldots,N$ and $m=1,\ldots,M$. We collect the channels from the entire RIS to BS antenna $m$ in $\vect{g}_m=[g_{m,1}, \ldots ,g_{m,N}]^{\Ttran}\in \mathbb{C}^{N}$ and the channels from RIS element $n$ to the entire BS array in $\vect{g}^{\prime}_n=[g_{1,n},\ldots ,g_{M,n}]^{\Ttran}\in \mathbb{C}^{M}$. We assume the channels $\vect{h}$ and $\vect{g}_m$ (and thus $\vect{g}^{\prime}_n$) are independent, and also assume that $\vect{g}_m\sim\CN(\vect{0}_N,\vect{R}_{\mathrm{g}})$ and  $\vect{g}^{\prime}_n\sim\CN(\vect{0}_M,\vect{R}_{\mathrm{g}^{\prime}})$ with $\vect{R}_{\mathrm{g}^{\prime}}\in \mathbb{C}^{M\times M}$ and $\vect{R}_{\mathrm{g}}\in \mathbb{C}^{N \times N}$ being respectively the receive and transmit correlation matrices for the BS and RIS, obtained through the Kronecker model \cite{Shiu2000a,Demir2021RIS}. The Kronecker model is built on the common assumption that the multipath scattering at the BS is uncorrelated with the scattering at the RIS side \cite{swindlehurst2022channel}.

\section{Pilot Transmission and Channel Estimation} \label{sec:pilot}

We neglect the direct link between the transmitter and the receiver to focus on the RIS phase-shift design during pilot transmission. We stress that no generality is lost under this assumption since the direct channel can be estimated separately by an orthogonal pilot scheme \cite{RISchannelEstimation_nested_knownBSRIS2,Demir2021RIS}. 
Unlike conventional communications, if a passive RIS is used, the BS must estimate the cascaded channel
\begin{align}
\vect{h}_m\triangleq \vect{h}\odot\vect{g}_m \in \mathbb{C}^N, \quad m=1,\ldots,M, 
\end{align}
in each coherence block to obtain complete channel state information, where $\odot$ denotes the Hadamard (element-wise) product.  This information is needed for the design of RIS phase-shifts and precoded signals during data transmission.

\subsection{Pilot Transmission}

To estimate $\vect{h}_m$, the UE can send a predefined pilot sequence $\boldsymbol{\psi} = [{\psi}(1), \ldots, {\psi}(\tau_p)]^{\Ttran}\in \mathbb{C}^{\tau_p}$ during $\tau_p \geq 1$ channel uses, where $\tau_p$ is the length of the pilot sequence. Its length cannot be longer than the number of channel uses in a coherence block. The classical RIS channel estimation papers require $\tau_p=N$, which might incur high overhead. In this paper, we consider channel estimation schemes that work under shorter pilots in a rich scattering environment.

The received signal at the RIS when ${\psi}(t)$ is transmitted (at discrete time $t$) is 
\begin{equation}
    {\bf x}(t) = {\bf h} \psi(t) + {\bf w}
    \end{equation}
where ${\bf w}\in \mathbb{C}^N$ is the EMI~\cite{Torres2021}, produced by any incoming electromagnetic waves that are independent of the pilot signal $\psi(t)$.\footnote{Here, we assume slowly varying EMI in line with the previous work \cite{long2023channel,khaleel2023electromagnetic}. Later in the numerical results, the case of fast-varying EMI will also be analyzed to quantify its impact.} We assume the EMI takes independent realizations in each coherence block according to the  distribution ${\bf w}\sim\CN(\vect{0}_N,\sigma^2_{\rm w}\vect{R}_{\rm w})$, where $\vect{R}_{\rm w}$ is defined in the same general matter as in~\eqref{eq:spatial-correlation} but with a different spatial scattering function 
$f_{\rm w}(\varphi,\theta)$:
\begin{equation} \label{eq:spatial-correlation-EMI}
\vect{R}_{\rm w} =  \iint_{-\pi/2}^{\pi/2} f_{\rm w}(\varphi,\theta) \vect{a}(\varphi,\theta) \vect{a}^{\Htran}(\varphi,\theta) d\theta d\varphi.
\end{equation}
The received signal at BS antenna $m$ is 
\begin{align}
    y_m(t) = \hspace{-0.5cm}\underbrace{\boldsymbol{\phi}^{\Ttran}(t)\vect{h}_m \psi(t)}_{\text{Pilot signal scattered by the RIS}} \!\!\!\!\!+ \!\!\!\!\!\overbrace{\boldsymbol{\phi}^{\Ttran}(t){\bf w}_m}^{\text{EMI scattered by the RIS}}\!\!\!\!\!\! +\!\!\!\! \underbrace{n_m(t)}_{\text{Noise at the BS}}
\end{align}
where 
\begin{equation}
{\bf w}_m = {\bf w}\odot\vect{g}_m 
\end{equation}
and $\boldsymbol{\phi}(t) =  [e^{-\imagunit\phi_{1}(t)},\ldots,e^{-\imagunit\phi_{N}(t)}]^{\Ttran}   \in \mathbb{C}^N$ is the vector collecting the RIS phase-shifts $\{\phi_{n}(t) : n=1,\ldots,N\}$ during the transmission of pilot $\psi(t)$. The noise $n_m(t) \sim \mathcal N(0,\sigma^2_{\rm n})$ accounts for any uncontrollable factor (e.g., of electromagnetic or hardware nature) disturbing the signal reception at the BS, except for the EMI scattered by the RIS whose statistics depend on the RIS configuration.

The received vector ${\bf y}_m = [y_m(1), \ldots, y_m(\tau_p)]^{\Ttran}$ at BS antenna $m$ during the $\tau_p$ channel uses takes the form
\begin{equation}
    {\bf y}_m = \diag({\boldsymbol \psi})\boldsymbol{\Phi}\vect{h}_m  + \boldsymbol{\Phi}{\bf w}_m + {\bf n}_m
\end{equation}
where $\vect{\Phi}\in \mathbb{C}^{\tau_p\times N}$ with $\left[\vect{\Phi}\right]_{t,n}=e^{-\imagunit\phi_{n}(t)}$ is the global RIS phase-shift matrix and ${\bf n}_m\sim \CN\left(\vect{0}_{\tau_p}, \sigma^2_{\rm n}\vect{I}_{\tau_p} \right)$. For simplicity, we assume that ${\psi}(t) = \sqrt{\rho}$ for all $t$ with $\rho\ge 0$ being the pilot transmit power.\footnote{This selection does not lead to loss of generality since the cumulative phase-shift of the pilot signals and the RIS can be represented by the entries of the RIS phase-shift matrix $\vect{\Phi}$.} With this selection and collecting the received signals for all the BS antennas, we obtain
\begin{equation} \label{eq:concat}
   \underbrace{ \begin{bmatrix}\vect{y}_1 \\ \vdots \\ \vect{y}_M \end{bmatrix}}_{\triangleq \vect{y}}= \sqrt{\rho} \vect{\Phi}_M\underbrace{ \begin{bmatrix}\vect{h}_1 \\ \vdots \\ \vect{h}_M \end{bmatrix}}_{\triangleq \vect{x}} +\,  \vect{\Phi}_M\underbrace{ \begin{bmatrix}\vect{w}_1 \\ \vdots \\ \vect{w}_M \end{bmatrix}}_{\triangleq \vect{z}} + \underbrace{ \begin{bmatrix}\vect{n}_1 \\ \vdots \\ \vect{n}_M \end{bmatrix}}_{\triangleq \vect{n}}
\end{equation}
where $ \vect{\Phi}_M \in \mathbb{C}^{M \tau_p \times MN}$ is given by
\begin{equation} \label{eq:Phi-tilde}
 \vect{\Phi}_M = \vect{I}_M \kron \vect{\Phi}.
\end{equation}

In this paper, we will consider different estimators of  $\vect{x}$ that are all based on the received signal $\vect{y}$ in \eqref{eq:concat} but require different prior statistical information. We will propose methods to select $\vect{\Phi}$ to optimize the estimation performance. The baseline is the classical LS estimator \cite{Bjornson2022a}, according to which
\begin{align}\label{eq:LS-estimate}
\widehat{\vect{x}}_{\rm LS} = \frac{1}{\sqrt{\rho}}  (\vect{\Phi}_M^{\Htran}\vect{\Phi}_M)^{-1}\vect{\Phi}_M^{\Htran} \vect{y}
\end{align}
that requires no prior information but needs $\tau_p \geq N$ to enable inversion of $\vect{\Phi}_M^{\Htran}\vect{\Phi}_M = \vect{I}_M \kron (\vect{\Phi}^{\Htran}\vect{\Phi})$.
Since RISs with very large $N$ are envisioned \cite{Bjornson2022a}, we are targeting other estimators that can allow us to have $\tau_p < N$. 
We will go through two such estimation methods in this section, under the assumption that $\vect{\Phi}$ has full rank but is otherwise arbitrarily selected.

\subsection{Linear Minimum Mean-Square Error Estimation}

From \eqref{eq:concat}, the observation vector for the estimate of $\vect{x}$ is
\begin{align} \label{eq:y-EMI} 
\vect{y} = \sqrt{\rho} \vect{\Phi}_M \vect{x} + \vect{\Phi}_M \vect{z}  +\vect{n}
\end{align}
where $\vect{z}$ and $\vect{n}$ are independent of each other and they are both uncorrelated with $\vect{x}$. 
If the first and second-order moments of $\vect{x}$ are known, we can use the LMMSE estimator.\footnote{Note that this estimator will not coincide with the (non-linear) MMSE estimator since $\vect{x}$ in \eqref{eq:concat} is not Gaussian distributed.}
Since the channel has zero mean, this is equivalent to knowing the full spatial correlation matrix $\vect{R}_{\rm x} =\mathbb{E}\{\vect{x}\vect{x}^{\Htran}\} \in \mathbb{C}^{MN\times MN}$, which is given by
\begin{align} \label{eq:Rx}
    \vect{R}_{\rm x} =\mathbb{E}\{\vect{x}\vect{x}^{\Htran}\} =\vect{R}_{{\rm g}^{\prime}} \kron \left(\vect{R}_{\rm h}\odot\vect{R}_{\rm g}\right)
\end{align}
where we have used the independence of $\vect{h}$ and $\{\vect{g}_m: m=1,\ldots,M\}$. 
Similarly, we obtain
\begin{align}\label{eq:Rz}
  \mathbb{E}\{\vect{z}\vect{z}^{\Htran}\} =\sigma^2_{\rm w}  \vect{R}_{\rm z} &=\sigma^2_{\rm w} \underbrace{\vect{R}_{{\rm g}^{\prime}} \kron \left(\vect{R}_{\rm w}\odot\vect{R}_{\rm g}\right)}_{= \vect{R}_{\rm z}}\end{align}
due to the independence of $\vect{w}$ and $\{\vect{g}_m : m=1,\ldots,M\}$.
Using $\vect{R}_{\rm x}$ and $\vect{R}_{\rm z}$, the LMMSE estimate of $\vect{x}$ is given by~\cite{Kay1993a}
\begin{align}\label{eq:LMMSE-estimate}
\widehat{\vect{x}}_{\rm LMMSE} = \sqrt{\rho}\vect{R}_{\rm x}\vect{\Phi}_M^{\Htran}\left(\rho\vect{\Phi}_M\vect{R}_{\rm x}\vect{\Phi}_M^{\Htran} + {\bf C}\right)^{-1}\vect{y}
\end{align}
with
\begin{align}
{\bf C} = \sigma^2_{\rm w}\vect{\Phi}_M\vect{R}_{\rm z}\vect{\Phi}_M^{\Htran}+\sigma^2_{\rm n}\vect{I}_{M\tau_p}.
\end{align}
The mean-square error (MSE) $\mathbb{E}\{||\widehat{\vect{x}}_{\rm LMMSE}  - \vect{x}||^2\}$ is~\cite{Kay1993a}
\begin{align} \label{eq:mse-lmmse}
 &   {\rm MSE}_{\rm LMMSE} = \tr(\vect{R}_{\rm x}) \nonumber\\
    &\quad -\rho\tr\left(\vect{R}_{\rm x}\vect{\Phi}_M^{\Htran}\left(\rho\vect{\Phi}_M\vect{R}_{\rm x}\vect{\Phi}_M^{\Htran}+{\bf C}\right)^{-1}\vect{\Phi}_M\vect{R}_{\rm x}\right).
\end{align}
Notice that we may rewrite $\widehat{\vect{x}}_{\rm LMMSE}$ as 
\begin{align}\label{eq:LMMSE-estimate-v1}
\widehat{\vect{x}}_{\rm LMMSE}=\frac{1}{\sqrt{\rho}}\vect{R}_{\rm x}\vect{\Phi}_M^{\Htran}\Bigg(&\vect{\Phi}_M\left(\vect{R}_{\rm x} + \frac{1}{\rm {SIR}} \vect{R}_{\rm z}\right) \vect{\Phi}_M^{\Htran} \nonumber\\
&+ \frac{1}{\rm {SNR}} \vect{I}_{M\tau_p}\Bigg)^{-1}\vect{y}
\end{align}
where we have defined the signal-to-noise ratio (SNR) in the pilot transmission phase as
\begin{align}
{\rm {SNR}} = \frac{\rho}{\sigma^2_{\rm n}}
\end{align}
and the signal-to-EMI ratio (SIR) as
\begin{align}
{\rm {SIR}} = \frac{\rho}{\sigma^2_{\rm w}}.
\end{align}
We notice that the EMI cannot be treated as additional colored noise (which can be removed using a classical whitening filter) since the EMI term $\vect{\Phi}_M \vect{z}$ in \eqref{eq:y-EMI} depends on the phase-shift matrix.
Hence, the selection of $\vect{\Phi}_M$ not only affects the desired signal but also changes the EMI characteristics. This calls for a separate analysis of the design of the RIS phase-shift pattern.

While the LS estimator in \eqref{eq:LS-estimate} requires $\tau_p=N$, the LMMSE estimator in \eqref{eq:LMMSE-estimate-v1} can be applied for any $\tau_p\geq 1$, but we should expect to get a better result for larger values of $\tau_p$ since we explore more channel dimensions.
The major practical drawback is that the $M^2N^2$ entries of the UE-specific correlation matrix $\vect{R}_{\rm x}$ must be known at the BS, which requires extensive measurements over many coherence blocks proportional to $M^2N^2$.
This may not be realistic in practical scenarios\footnote{By contrast, the EMI correlation matrix $\sigma^2_{\rm w}  \vect{R}_{\rm z}$ is the same for any UE connecting to the BS. Hence, it can possibly be assumed known. It can be estimated by sensing the wireless channel when the UEs are not transmitting pilots.} and will be addressed in the next section.

\subsection{Reduced-Subspace Least Squares Estimation}

We want to alleviate the need to have UE-specific statistical information, but without considering the LS estimator, which is known to provide overly conservative results when considering large dense arrays \cite{demir2022channel}. One structural property that can be exploited is the array geometry, which remains the same regardless of which UE is connected.
In this section, we will propose an RS-LS channel estimator for RIS-aided communications, inspired by our previous RS-LS estimator for holographic massive MIMO in \cite{demir2022channel}.

\begin{figure*}
 \begin{align} \label{eq:eigdecomp}
 \vect{R}_{\rm x}=\vect{U}_{\rm x}\vect{D}_{\rm x}\vect{U}_{\rm x}^{\Htran}  = \underbrace{\begin{bmatrix} \vect{U}_{\rm x,1}  & \vect{U}_{\rm x,2} \end{bmatrix}}_{\vect{U}_{\rm x}} \underbrace{\begin{bmatrix} \vect{D}_{\rm x,1} & \vect{0}_{r_{\rm x}\times (MN-r_{\rm x})} \\
 \vect{0}_{(MN-r_{\rm x})\times r_{\rm x}} & \vect{0}_{(MN-r_{\rm x})\times(MN-r_{\rm x})}\end{bmatrix}}_{\vect{D}_{\rm x}}  \underbrace{\begin{bmatrix} \vect{U}_{\rm x,1}^{\Htran} \\
 \vect{U}_{\rm x,2}^{\Htran}\end{bmatrix}}_{\vect{U}_{\rm x}^{\Htran}}.
 \end{align}
 \vspace{-6mm}
 \hrulefill
 \end{figure*}

Let the eigendecomposition of  $\vect{R}_{\rm x}$ be given in \eqref{eq:eigdecomp} at the top of the next page, where $\vect{U}_{\rm x}\in \mathbb{C}^{MN \times MN}$ is the matrix containing the orthonormal eigenvectors of $\vect{R}_{\rm x}$ and $\vect{D}_{\rm x}\in\mathbb{C}^{MN \times MN}$ is the diagonal matrix with the real-valued eigenvalues appearing in descending order. The rank of $\vect{R}_{\rm x}$ is denoted $r_{\rm x}=\rank(\vect{R}_{\rm x})$. The portions  $\vect{U}_{\rm x,1}\in \mathbb{C}^{MN\times r_{\rm x}}$ and $\vect{U}_{\rm x,2}\in \mathbb{C}^{MN \times (MN-r_{\rm x})}$ of $\vect{U}_{\rm x}$ contain the eigenvectors from $\vect{R}_{\rm x}$ corresponding to the $r_{\rm x}$ positive and $(MN-r_{\rm x})$ zero-valued eigenvalues, respectively. The diagonal matrix $\vect{D}_{\rm x,1}\in \mathbb{C}^{r_{\rm x}\times r_{\rm x}}$ contains all the positive eigenvalues. 
 
Any channel realization $\vect{x}$ can be expressed as $\vect{U}_{\rm x,1}\vect{a}$ for some vector $\vect{a} \in \mathbb{C}^{r_{\rm x}}$. In fact, the elements of $\vect{a}$ are independent but have different variances determined by $\vect{D}_{\rm x,1}$. A major problem is that $\vect{D}_{\rm x,1}$ differs between UEs, and we aim to design an estimator that does not require such UE-specific information. Therefore, we want to apply LS estimation, but only within the subspace spanned by $\vect{U}_{\rm x,1}$.
  This logic leads to the following RS-LS estimation of $\vect{x}$:
  
\begin{enumerate}
    \item Compute the LS estimate of $\vect{a}$ in the subspace spanned by $\vect{U}_{\rm x,1}$ by multiplying $\vect{y}$ by the pseudoinverse of $\vect{\Phi}_M \vect{U}_{\rm x,1}$ and then dividing by $\sqrt{\rho}$;
    \item Multiply the estimate by $\vect{U}_{\rm x,1}$ to get back to the original $MN$ dimensional space.
\end{enumerate}
More precisely, the RS-LS estimate of $\vect{x}$ is obtained as
\begin{align} \label{eq:RS-LS-estimate}
    \widehat{\vect{x}}_{\rm RS-LS} = \frac{\vect{U}_{\rm x,1}\left(\vect{U}_{\rm x,1}^{\Htran}\vect{\Phi}_M^{\Htran}\vect{\Phi}_M\vect{U}_{\rm x,1}\right)^{-1}\vect{U}_{\rm x,1}^{\Htran}\vect{\Phi}_M^{\Htran}\vect{y}}{\sqrt{\rho}},
\end{align}
where the pseudo-inversion requires $M\tau_p\geq r_{\rm x}$. 
Hence, the RS-LS estimation method relaxes the original requirement $\tau_p \geq N$ for LS estimation to $\tau_p\geq r_{\rm x}/M$, and also removes noise and EMI from unused channel dimensions when $r_{\rm x}<MN$. Using standard Kronecker product identities, we can express $\vect{U}_{\rm x,1}$ as $\vect{U}_{\rm x,1}=\vect{U}_{{\rm g}^{\prime},1}\kron\vect{U}_{{\rm h g},1}$ where $\vect{U}_{{\rm g}^{\prime},1}\in\mathbb{C}^{M\times r_{{\rm g}^{\prime}}}$ and $\vect{U}_{{\rm h  g},1}\in\mathbb{C}^{N\times r_{\rm h  g}}$ consist of the orthonormal eigenvectors of the correlation matrices
$\vect{R}_{{\rm g}^{\prime}}$ and $\vect{R}_{\rm h}\odot\vect{R}_{\rm g}$, respectively, corresponding to the non-zero eigenvalues. These matrices represent the spatial correlation characteristics at the BS and RIS sides, respectively. We
can then compute the rank of $\vect{R}_{\rm x}$ as $r_{\rm x}=r_{{\rm g}^{\prime}}r_{\rm hg}$, where  $r_{{\rm g}^{\prime}}=\rank(\vect{R}_{{\rm g}^{\prime}})$ and $r_{\rm h g}=\rank(\vect{R}_{\rm h}\odot\vect{R}_{\rm g})$.

The RS-LS estimator above still uses UE-specific statistical information, since the column space of $\vect{U}_{\rm x,1}$ can vary between UEs.
However, we could replace $\vect{U}_{\rm x,1}$ with a matrix representing the range of all plausible correlation matrices. 
The following lemma provides a condition for when the span of a cascaded spatial correlation matrix is within the span of another correlation matrix. It is an extension of \cite[Lem.~3]{demir2022channel}, which considered scenarios without RIS.

\begin{lemma} \label{lemma:span} Let $\overline{\vect{R}}_{\rm x }=\overline{\vect{R}}_{{\rm g}^{\prime}} \kron \left(\overline{\vect{R}}_{\rm h}\odot\overline{\vect{R}}_{\rm g}\right)$ and $\vect{R}_{\rm x}=\vect{R}_{{\rm g}^{\prime}} \kron \left(\vect{R}_{\rm h}\odot\vect{R}_{\rm g}\right)$ be two spatial correlation matrices for the cascaded channel obtained using the same RIS and BS array geometries. Let the spatial scattering functions corresponding to the correlation matrices $\overline{\vect{R}}_{i}$ and $\vect{R}_i$ according to the correlated fading model in \eqref{eq:spatial-correlation} be denoted by $\overline{f}_{i}(\varphi,\theta)$ and $f_i(\varphi,\theta)$, respectively, for $i\in \{{\rm g}^{\prime},{\rm h}, {\rm g}\}$, $\varphi\in[-\pi/2,\pi/2]$ and $\theta\in[-\pi/2,\pi/2]$. We assume that the spatial scattering functions are either continuous at each point on its domain or contain Dirac delta functions.

If the domain of  $\overline{f}_i(\varphi,\theta)$ for which $\overline{f}_i(\varphi,\theta)>0$ contains the domain of  $f_i(\varphi,\theta)$ for which $f_i(\varphi,\theta)>0$, for $i\in\{{\rm g}^{\prime},{\rm h},{\rm g}\}$, then the subspace spanned by the columns of $\overline{\vect{R}}_{\rm x}$ contains the subspace spanned by the columns of $\vect{R}_{\rm x}$. 
\end{lemma}

\begin{IEEEproof}
The proof extends the proof of \cite[Lem.~3]{demir2022channel} by defining new spatial scattering functions $f_{{\rm g}^{\prime}}(\varphi_1,\theta_1)f_{\rm h}(\varphi_2,\theta_2)f_{\rm g}(\varphi_3,\theta_3)$ and array response vectors $\vect{a}_{\rm BS}(\varphi_1,\theta_1)\kron\left(\vect{a}_{\rm RIS}(\varphi_2,\theta_2)\odot\vect{a}_{\rm RIS}(\varphi_3,\theta_3)\right)$ for the UPAs in terms of $(\varphi_1,\theta_1,\varphi_2,\theta_2,\varphi_3,\theta_3)$ on the six-dimensional angular domain. 
\end{IEEEproof}

According to Lemma~\ref{lemma:span}, a channel with non-zero spatial scattering functions for all angles will have correlation matrices that span all other correlation matrices.
A primary example is
\begin{align} \label{eq:conservative-Rx}
\overline{\vect{R}}_{\rm x}=\vect{R}_{\rm BS, iso}\kron\left(\vect{R}_{\rm RIS, iso}\odot\vect{R}_{\rm RIS, iso}\right) 
\end{align}
where $\vect{R}_{\rm BS, iso}$  and $\vect{R}_{\rm RIS, iso}$ are the spatial correlation matrices for isotropic scattering from \eqref{R-iso} (according to the BS or RIS array geometry, respectively).

Using this correlation matrix together with the RS-LS estimator, we ensure that all physically plausible channel dimensions are considered by the estimator, without requiring any UE-specific prior information; only the array geometries are exploited.
This leads to the \emph{conservative} RS-LS estimator given in \eqref{eq:RS-LS-estimate-approx} at the top of the next page
\begin{figure*}
\begin{align} \label{eq:RS-LS-estimate-approx}
    \widehat{\vect{x}}_{\rm RS-LS}^{\rm conserv} &= \frac{\overline{\vect{U}}_{\rm x,1}\left(\overline{\vect{U}}_{\rm x,1}^{\Htran}\vect{\Phi}_M^{\Htran}\vect{\Phi}_M\overline{\vect{U}}_{\rm x,1}\right)^{-1}\overline{\vect{U}}_{\rm x,1}^{\Htran}\vect{\Phi}_M^{\Htran}\vect{y}}{\sqrt{\rho}} =  \underbrace{\overline{\vect{U}}_{\rm x,1}\vect{a}}_{=\vect{x}}\;\; +\;\;\frac{\overline{\vect{U}}_{\rm x,1}\left(\overline{\vect{U}}_{\rm x,1}^{\Htran}\vect{\Phi}_M^{\Htran}\vect{\Phi}_M\overline{\vect{U}}_{\rm x,1}\right)^{-1}\overline{\vect{U}}_{\rm x,1}^{\Htran}\vect{\Phi}_M^{\Htran}\left(\vect{\Phi}_M \vect{z}  +\vect{n}\right)}{\sqrt{\rho}}.
\end{align}
\vspace{-6mm}
\hrulefill
\end{figure*}
where the columns $\overline{\vect{U}}_{\rm x,1}\in \mathbb{C}^{MN \times \overline{r}_{\rm x}}$ 
are the orthonormal eigenvectors corresponding to the  $ \overline{r}_{\rm x}$ non-zero eigenvalues of $\overline{\vect{R}}_{\rm x}$ in \eqref{eq:conservative-Rx}.
To apply this estimator, we must choose $\tau_p$ such that $\overline{r}_{\rm x}\leq M \tau_p$ so that the pseudo-inverse is well defined.
The second term in~\eqref{eq:RS-LS-estimate-approx} represents the estimation error due to EMI $\vect{z}$ and noise $\vect{n}$, and this term depends on the phase-shift matrix $\vect{\Phi}_M$.
Thus, the choice of the phase-shift matrix determines the channel estimation accuracy.
The corresponding MSE $\mathbb{E}\{||\widehat{\vect{x}}_{\rm RS-LS}^{\rm conserv}   - \vect{x}||^2\}$ takes the form given in \eqref{eq:mse-rsls} at the top of the next page.
\begin{figure*}
\begin{align} \label{eq:mse-rsls}
    {\rm MSE}_{\rm RS-LS}^{\rm conserv}&= \frac{\tr\left(\overline{\vect{U}}_{\rm x,1}\left(\overline{\vect{U}}_{\rm x,1}^{\Htran}\vect{\Phi}_M^{\Htran}\vect{\Phi}_M\overline{\vect{U}}_{\rm x,1}\right)^{-1}\overline{\vect{U}}_{\rm x,1}^{\Htran}\right)}{{\rm {SNR}}} \nonumber\\
    &+\frac{\tr\left(\overline{\vect{U}}_{\rm x,1}\left(\overline{\vect{U}}_{\rm x,1}^{\Htran}\vect{\Phi}_M^{\Htran}\vect{\Phi}_M\overline{\vect{U}}_{\rm x,1}\right)^{-1}\overline{\vect{U}}_{\rm x,1}^{\Htran}\vect{\Phi}_M^{\Htran}\vect{\Phi}_M\vect{R}_{z}\vect{\Phi}_M^{\Htran}\vect{\Phi}_M\overline{\vect{U}}_{\rm x,1}\left(\overline{\vect{U}}_{\rm x,1}^{\Htran}\vect{\Phi}_M^{\Htran}\vect{\Phi}_M\overline{\vect{U}}_{\rm x,1}\right)^{-1}\overline{\vect{U}}_{\rm x,1}^{\Htran}\right)}{{\rm {SIR}}} \nonumber \\
    &\hspace{-20mm}=\frac{\tr\left(\left(\overline{\vect{U}}_{\rm x,1}^{\Htran}\vect{\Phi}_M^{\Htran}\vect{\Phi}_M\overline{\vect{U}}_{\rm x,1}\right)^{-1}\right)}{{\rm SNR}} +\frac{\tr\left(\left(\overline{\vect{U}}_{\rm x,1}^{\Htran}\vect{\Phi}_M^{\Htran}\vect{\Phi}_M\overline{\vect{U}}_{\rm x,1}\right)^{-1}\overline{\vect{U}}_{\rm x,1}^{\Htran}\vect{\Phi}_M^{\Htran}\vect{\Phi}_M\vect{R}_{z}\vect{\Phi}_M^{\Htran}\vect{\Phi}_M\overline{\vect{U}}_{\rm x,1}\left(\overline{\vect{U}}_{\rm x,1}^{\Htran}\vect{\Phi}_M^{\Htran}\vect{\Phi}_M\overline{\vect{U}}_{\rm x,1}\right)^{-1}\right)}{{\rm SIR}}.
\end{align}
\vspace{-6mm}
\hrulefill
\end{figure*}

\begin{remark}
While obtaining the correlation matrices between the RIS and the UEs is challenging in practical scenarios, the correlation matrix between the BS and the RIS can be known given their fixed and elevated locations. In such cases, it is possible to design a channel estimation scheme using LMMSE and RS-LS for different segments of the RIS channel. For example, the BS-RIS and RIS-UE channels can be estimated iteratively using LMMSE and RS-LS, respectively.
\end{remark}

\begin{remark}
Although we assume spatially correlated Rayleigh fading for the channel distributions, the LMMSE and RS-LS estimators can also be applied in scenarios with a line-of-sight (LOS) component, such as under Rician fading. The LMMSE estimator can be utilized with the spatial correlation matrix of the overall channel, which includes both LOS and non-line-of-sight (NLOS) components. The RS-LS estimator can be used as presented here because the LOS component does not alter the subspace induced by the array geometry.
\end{remark}

The computational complexity of the LMMSE estimator involves taking the inverse of a $M\tau_p \times M\tau_p$ matrix and performing matrix multiplications of sizes $MN \times MN$ and $MN \times M\tau_p$. The total complexity is given by $\mathcal{O}(M^3(2N^2\tau_p+3N\tau_p^2+\tau_p^3/3))$ \cite[Lem.~B.1 and B.2]{massivemimobook}. For the RS-LS estimator, the complexity involves inverting a matrix of size $\overline{r}_{\rm x}\times \overline{r}_{\rm x}$ and multiplying matrices of sizes $\overline{r}_{\rm x}\times M\tau_p$ and $M\tau_p \times MN$. The total complexity is  $\mathcal{O}(2M^2N\tau_p\overline{r}_{\rm x}+\overline{r}_{\rm x}^3/3+2M\tau_p \overline{r}_{\rm x}^2)$. However, once the matrices that remain constant over many coherence blocks are computed, channel estimation can be achieved with a complexity of at most 
$\mathcal{O}(M^2N\tau_p)$ in each coherence block. This is lower than iterative methods, such as the algorithm presented in \cite{huang2022semi}, especially when $N>M$.

We now have two different MSE expressions depending on the phase-shift matrix $\vect{\Phi}$:  
${\rm MSE}_{\rm LMMSE}$ in \eqref{eq:mse-lmmse} for the LMMSE estimator and ${\rm MSE}_{\rm RS-LS}^{\rm conserv}$ in \eqref{eq:mse-rsls} for the conservative RS-LS estimator.
In the next two sections, we will minimize these MSEs with respect to the phase-shift matrix $\vect{\Phi}$.

\section{Optimizing the RIS Configuration for LMMSE Estimation} \label{sec:optimization-LMMSE}

The aim of this section is to find the $\vect{\Phi}$ that minimizes ${\rm MSE}_{\rm LMMSE}$ in \eqref{eq:mse-lmmse} under the constraints
 $|\left[\vect{\Phi}\right]_{t,n}|=1$, for $t=1,\ldots,\tau_p$ and $n=1,\ldots,N$.  As a first step, we consider the relaxed problem given in \eqref{eq:lmmse-opt} at the top of the next page,
 \begin{figure*}
\begin{align}
& \underset{\tr\left(\vect{\Phi}^{\Htran}\vect{\Phi}\right)\leq N\tau_p}{\mathacr{maximize}} \ \ \tr\left(\vect{R}_{\rm x}\vect{\Phi}_M^{\Htran}\left(\vect{\Phi}_M\vect{R}_{\rm x}\vect{\Phi}_M^{\Htran}+\frac{1}{\rm{SIR}}\vect{\Phi}_M\vect{R}_{\rm z}\vect{\Phi}_M^{\Htran}+\frac{1}{\rm{SNR}}\vect{I}_{M\tau_p}\right)^{-1}\vect{\Phi}_M\vect{R}_{\rm x}\right).   \label{eq:lmmse-opt}
\end{align}
\vspace{-6mm}
\hrulefill
\end{figure*}
where the non-convex unit modulus constraints have been replaced by a Frobenius norm constraint on the matrix $\vect{\Phi}$ (and we also removed the constant part of the objective function). After obtaining the optimal solution to the approximate problem in \eqref{eq:lmmse-opt}, we will project it to have $|\left[\vect{\Phi}\right]_{t,n}|=1$, for $t=1,\ldots,\tau_p$ and $n=1,\ldots,N$. 
We will first solve \eqref{eq:lmmse-opt} in the simpler case without EMI, i.e., $\frac{1}{\rm{SIR}}=0$ and then the general case with EMI.

\subsection{Optimal Structure without EMI}
Minimizing the MSE in \eqref{eq:mse-lmmse} is equivalent to maximizing the second term. When the EMI is absent (or negligible), this problem reduces to 
\begin{align}
& \underset{\tr\left(\vect{\Phi}^{\Htran}\vect{\Phi}\right)\leq N\tau_p}{\mathacr{maximize}} \nonumber\\
&\tr\left(\vect{R}_{\rm x}\vect{\Phi}_M^{\Htran}\left(\vect{\Phi}_M\vect{R}_{\rm x}\vect{\Phi}_M^{\Htran}+\frac{1}{\rm{SNR}}\vect{I}_{M\tau_p}\right)^{-1}\vect{\Phi}_M\vect{R}_{\rm x}\right),   \label{eq:lmmse-opt-no-EMI}
\end{align}
where the constant terms are omitted and the unit-modulus constraints are relaxed. The above problem and its specific variants have been solved in a different context, i.e., training signal design for classical MIMO with correlated channels
\cite{Kotecha2004a,Liu2007a,Bjornson2010a}. In the following lemma, we outline the optimal structure of $\vect{\Phi}$ solving~\eqref{eq:lmmse-opt-no-EMI}.

\begin{lemma} \label{lemma2}
 Let the eigendecomposition of  $\vect{R}_{{\rm g}^{\prime}}$ and $\vect{R}_{\rm h}\odot\vect{R}_{\rm g}$  be given by $\vect{R}_{{\rm g}^{\prime}}=\vect{U}_{{\rm g}^{\prime}}\vect{D}_{{\rm g}^{\prime}}\vect{U}_{{\rm g}^{\prime}}^{\Htran}$ and $\vect{R}_{\rm h}\odot\vect{R}_{\rm g}=\vect{U}_{\rm hg}\vect{D}_{\rm hg}\vect{U}_{\rm hg}^{\Htran}$, respectively. $\vect{D}_{{\rm g}^{\prime}}\in\mathbb{C}^{M \times M}$ and  $\vect{D}_{\rm hg}\in\mathbb{C}^{N \times N}$  are the diagonal matrices with the real non-negative eigenvalues appearing on the diagonals in descending order, which are $d_{{\rm g}^{\prime}, 1}\geq \ldots \geq d_{{\rm g}^{\prime},M}\geq 0$ and $d_{{\rm hg}, 1}\geq \ldots \geq d_{{\rm hg},N}\geq 0$, respectively.  The solution to~\eqref{eq:lmmse-opt-no-EMI} has the structure
 \begin{align}\vect{\Phi}^{\star}=\vect{S}_{\rm \Phi}\vect{\Lambda}_{\rm \Phi}\vect{U}_{\rm hg}^{\Htran} 
 \end{align}
 where $\vect{S}_{\rm \Phi}\in \mathbb{C}^{\tau_p\times\tau_p}$ is an arbitrary unitary matrix and the singular-value matrix $\vect{\Lambda}_{\rm \Phi} = \diag\{\lambda_{{\rm \Phi},1},\ldots, \lambda_{{\rm \Phi},N}\}$ has entries obtained by solving
 \begin{align}
  & \underset{\sum_{i=1}^{\tau_p} \lambda_{{\rm \Phi}, i}^2\leq N\tau_p}{\mathacr{minimize}} \ \  \sum_{i=1}^{\tau_p}\sum_{m=1}^M \frac{d_{{\rm g}^{\prime},m}d_{{\rm hg},i} }{ d_{{\rm g}^{\prime},m}d_{{\rm hg},i}\lambda_{{\rm \Phi},i}^2+\frac{1}{\rm{SNR}}}. \label{eq:objective3}
 \end{align}
\end{lemma}

\begin{IEEEproof}
The proof follows from \cite{Kotecha2004a} by noting that the left singular vectors do not affect the objective value.
\end{IEEEproof}
Note that the optimization problem in \eqref{eq:objective3} can be cast in a convex programming form by defining $\lambda_{{\rm \Phi},i}^2$ as the new optimization variables. The solution $\vect{\Phi}^{\star}$ from Lemma~\ref{lemma2} solves the relaxed problem in \eqref{eq:lmmse-opt-no-EMI} but the RIS only supports matrices with unit-modulus entries. Hence, we propose using the phase-shift matrix  $\vect{\Phi}=e^{\imagunit\angle{ \vect{\Phi}^{\star}}}$ where each entry is replaced by a complex exponential having the same phase.

\begin{corollary} \label{cor:M1}
When there is a single BS antenna (i.e., $M=1$), the singular values in $\vect{\Lambda}_{\rm \Phi}$ are
\begin{align} \label{eq:special-single}
    \lambda_{{\rm \Phi},i} = \sqrt{\max\left(0,\frac{1}{\sqrt{\mu}}-\frac{{\rm SNR}^{-1}}{d_{{\rm hg},i}}\right)}, \quad i=1,\ldots,\tau_p
 \end{align}
 where the Lagrange multiplier $\mu>0$ satisfies the constraint
 \begin{align} \label{eq:special-single2}
     \sum_{i=1}^{\tau_p}\max\left(0,\frac{1}{\sqrt{\mu}}-\frac{{\rm SNR}^{-1}}{ d_{{\rm hg},i}}\right)=N\tau_p.
 \end{align}

  \end{corollary}
  The solution for $M=1$ in Corollary~\ref{cor:M1} has the classical water-filling structure.
  In this special case,  the spatial correlation matrix $\vect{R}_{\rm x}$ in \eqref{eq:Rx} becomes $\vect{R}_{\rm h}\odot\vect{R}_{\rm g}$. From the water-filling structure in \eqref{eq:special-single}, it is seen that when the eigenvalues $d_{{\rm hg},i}$ of this matrix are similar (i.e., when the spatial correlation is low), the singular values of the phase-shift matrix are also similar. Since the matrix $\vect{S}_{\rm \Phi}$ can be chosen arbitrarily (see Lemma~\ref{lemma2}), as long as the phase-shift matrix $\vect{\Phi}$ is semi-unitary, we obtain the optimal performance. The selection of $\vect{\Phi}$ according to the spatial correlation matrices plays an important role when the eigenvalues $d_{{\rm hg},i}$ are very different (i.e., high spatial correlation). In such a scenario, more weight is given to the right singular vectors (eigenvectors of $\vect{R}_{\rm h}\odot\vect{R}_{\rm g}$) with higher eigenvalues in the construction of the phase-shift matrix $\vect{\Phi}$. This implies a greater emphasis on estimating the channel in the spatial directions where it is likely to be strong.
  Similar observations can be made for $M>1$, but then the eigenvalues of $\vect{R}_{\rm x}$ are $d_{{\rm g}^{\prime},m}d_{{\rm hg},i}$ as seen from \eqref{eq:objective3}. Moreover, there is a more complicated relationship between these eigenvalues and the singular values of the phase-shift matrix, since the same singular value $\lambda_{{\rm \Phi},i}$ appears in multiple terms, each related to a different BS antenna.
 
 \subsection{Optimal Structure with EMI}
 We will now optimize the phase-shift matrix in the case with EMI, which calls for separate derivations
 that differ substantially from previous works. We first define the matrix $\vect{B}\in\mathbb{C}^{MN\times MN}$ as
\begin{align} \label{eq:B}
    \vect{B} &= \vect{R}_{\rm x}+\frac{1}{\rm{SIR}}\vect{R}_{\rm z}\nonumber\\
    &=\vect{R}_{{\rm g}^{\prime}} \kron\underbrace{\left(\left(\vect{R}_{\rm h}\odot\vect{R}_{\rm g}\right)+\frac{1}{\rm{SIR}}\left(\vect{R}_{\rm w}\odot\vect{R}_{\rm g}\right)\right)}_{\triangleq\; \ddot{\vect{B}}\in\mathbb{C}^{N\times N} }
\end{align}
where we have used \eqref{eq:Rx} and \eqref{eq:Rz} together with the linearity property of the Kronecker product. The compact eigendecomposition of  $\vect{R}_{{\rm g}^{\prime}}$ is given by  $\vect{R}_{{\rm g}^{\prime}}=\vect{U}_{{\rm g}^{\prime},1}\vect{D}_{{\rm g}^{\prime},1}\vect{U}_{{\rm g}^{\prime},1}^{\Htran}$ where $\vect{U}_{{\rm g}^{\prime},1}\in\mathbb{C}^{M\times r_{{\rm g}^{\prime}}}$  consists of the orthonormal eigenvectors  corresponding to $r_{{\rm g}^{\prime}}=\rank(\vect{R}_{{\rm g}^{\prime}})$ positive eigenvalues. These eigenvalues appear on the entries of the diagonal matrix $\vect{D}_{{\rm g}^{\prime},1}\in \mathbb{C}^{r_{{\rm g}^{\prime}}\times r_{{\rm g}^{\prime}}}$ in descending order. In a similar manner, let the eigendecomposition of $\ddot{\vect{B}}$ in \eqref{eq:B} be given as $\ddot{\vect{B}}=\ddot{\vect{U}}_{\rm B,1}\ddot{\vect{D}}_{\rm B,1}\ddot{\vect{U}}_{\rm B,1}^{\Htran}$ where $\ddot{\vect{U}}_{\rm B,1}\in\mathbb{C}^{N\times \ddot{r}_{\rm B}}$ and $\ddot{\vect{D}}_{\rm B,1}\in\mathbb{C}^{\ddot{r}_{\rm B}\times \ddot{r}_{\rm B}}$ with $\ddot{r}_{\rm B}=\rank(\ddot{\vect{B}})$. By using standard Kronecker product identities, the eigenvalue decomposition of $\vect{B}$ in \eqref{eq:B} can be expressed as
\begin{align} \label{eq:B-eigenvalue}
    \vect{B} &= \vect{U}_{\rm B,1}\vect{D}_{\rm B,1}\vect{U}_{\rm B,1}^{\Htran} \nonumber\\ &= \underbrace {\left(\vect{U}_{{\rm g}^{\prime},1}\kron \ddot{\vect{U}}_{\rm B,1}\right)}_{\vect{U}_{\rm B,1}\in\mathbb{C}^{MN\times r_{\rm B}}}\underbrace {\left(\vect{D}_{{\rm g}^{\prime},1}\kron \ddot{\vect{D}}_{\rm B,1}\right)}_{\vect{D}_{\rm B,1}\in\mathbb{C}^{r_{\rm B}\times r_{\rm B}}}\underbrace {\left(\vect{U}_{{\rm g}^{\prime},1}^{\Htran}\kron \ddot{\vect{U}}^{\Htran}_{\rm B,1}\right)}_{\vect{U}_{\rm B,1}^{\Htran}\in\mathbb{C}^{r_{\rm B}\times MN}}
\end{align}
where $r_{\rm B}=\rank(\vect{B})=r_{{\rm g}^{\prime}}\ddot{r}_{\rm B}$. The following lemma will be useful.

\begin{lemma} \label{lemma:useful1}
For any given correlation matrices $\vect{R}_{\rm x}\in \mathbb{C}^{MN\times MN}$ and $\vect{B}=\vect{R}_{\rm x}+\frac{1}{\rm{SIR}}\vect{R}_{\rm z}\in \mathbb{C}^{MN \times MN}$ in \eqref{eq:B}, one can always find a matrix $\vect{F}\in \mathbb{C}^{r_{\rm B}\times r_{\rm x}}$ such that
\begin{align} \label{eq:Rx-lemma}
    \vect{R}_{\rm x} = \vect{U}_{\rm B,1}\vect{D}_{\rm B,1}^{\frac{1}{2}}\vect{F}\vect{F}^{\Htran}\vect{D}_{\rm B,1}^{\frac{1}{2}}\vect{U}_{\rm B,1}^{\Htran}
\end{align}
where $r_{\rm x}=\rank(\vect{R}_{\rm x})\leq r_{\rm B}$.
\end{lemma}
\begin{IEEEproof} The proof is given in Appendix~\ref{appendix-lemma-useful1}.
\end{IEEEproof}
 
 Now, let us define the positive semi-definite matrix
\begin{align} \label{eq:A}
    \vect{A}\triangleq\vect{\Phi}_M\vect{B}\vect{\Phi}_M^{\Htran}=\vect{\Phi}_M\vect{U}_{\rm B,1}\vect{D}_{\rm B,1}\vect{U}_{\rm B,1}^{\Htran}\vect{\Phi}_M^{\Htran} 
\end{align}
and denote its eigenvalue decomposition as in \eqref{eq:eigdecompA} at the top of the next page,
\begin{figure*}
 \begin{align} \label{eq:eigdecompA}
 \vect{A}=\vect{U}_{\rm A}\vect{D}_{\rm A}\vect{U}_{\rm A}^{\Htran}  = \underbrace{\begin{bmatrix} \vect{U}_{\rm A,1}  & \vect{U}_{\rm A,2} \end{bmatrix}}_{\vect{U}_{\rm A}} \underbrace{\begin{bmatrix} \vect{D}_{\rm A,1} & \vect{0}_{r_{\rm B}\times (M\tau_p-r_{\rm B})} \\
 \vect{0}_{(M\tau_p-r_{\rm B})\times r_{\rm B}} & \vect{0}_{(M\tau_p-r_{\rm B})\times(M\tau_p-r_{\rm B})}\end{bmatrix}}_{\vect{D}_{\rm A}}  \underbrace{\begin{bmatrix} \vect{U}_{\rm A,1}^{\Htran} \\
 \vect{U}_{\rm A,2}^{\Htran}\end{bmatrix}}_{\vect{U}_{\rm A}^{\Htran}}.
 \end{align}
 \vspace{-6mm}
 \hrulefill
 \end{figure*}
 where the unitary matrix $\vect{U}_{\rm A}\in \mathbb{C}^{M\tau_p \times M\tau_p}$ consists of the eigenvectors of $\vect{A}$ and $\vect{D}_{\rm A}\in\mathbb{C}^{M\tau_p \times M\tau_p}$ is the diagonal matrix with the non-negative eigenvalues in descending order. We know that the rank of $\vect{A}$ cannot exceed the rank of $\vect{B}=\vect{U}_{\rm B,1}\vect{D}_{\rm B,1}\vect{U}_{\rm B,1}^{\Htran}$ from \eqref{eq:A}. The parts $\vect{U}_{\rm A,1}\in \mathbb{C}^{M\tau_p\times r_{\rm B}}$ and $\vect{U}_{\rm A, 2}\in \mathbb{C}^{M\tau_p \times (M\tau_p-r_{\rm B})}$ of $\vect{U}_{\rm A}$ are the matrices whose columns are the orthonormal eigenvectors of $\vect{A}$ corresponding to the largest $r_{\rm B}$ and the remaining $(M\tau_p-r_{\rm B})$ zero eigenvalues, respectively.  
From \eqref{eq:trace1app} in Appendix~\ref{appendix-equation-steps}, the objective function in \eqref{eq:lmmse-opt} can be written as
\begin{align}
&   \sum_{i=1}^{r_{\rm B}}\frac{g_{i,i}d_{{\rm A},i}}{d_{{\rm A},i}+\frac{1}{\rm{SNR}}} \label{eq:trace1}
\end{align}
where $g_{i,i}\geq0$ and $d_{{\rm A},i}\geq 0$ denote the $(i,i)$th entry of $\vect{G}\triangleq\vect{F}\vect{F}^{\Htran}\vect{D}_{\rm B,1}\vect{F}\vect{F}^{\Htran}\in \mathbb{C}^{r_{\rm B} \times r_{\rm B}}$  and $\vect{D}_{\rm A,1}$, respectively. We notice that maximum value of \eqref{eq:trace1} is monotonically increasing with $\sum_{i=1}^{r_{\rm B}}d_{{\rm A},i}=\tr(\vect{D}_{\rm A,1})=\tr(\vect{A})$. By using \eqref{eq:A} and recalling \eqref{eq:Phi-tilde} and \eqref{eq:B-eigenvalue}, we obtain
\begin{align} \label{eq:traceA}
    \tr(\vect{A}) &=  \tr\left(\vect{\Phi}_M\vect{U}_{\rm B,1}\vect{D}_{\rm B,1}\vect{U}_{\rm B,1}^{\Htran}\vect{\Phi}_M^{\Htran}\right) \nonumber \\
    &= \tr\Big(\left(\vect{I}_{M}\kron\vect{\Phi}\right) \left(\vect{U}_{{\rm g}^{\prime},1}\kron \ddot{\vect{U}}_{\rm B,1}\right) \nonumber\\
    &\quad \times\left(\vect{D}_{{\rm g}^{\prime},1}\kron \ddot{\vect{D}}_{\rm B,1}\right)\left(\vect{U}_{{\rm g}^{\prime},1}^{\Htran}\kron \ddot{\vect{U}}^{\Htran}_{\rm B,1}\right)\left(\vect{I}_{M}\kron\vect{\Phi}^{\Htran}\right)\Big) \nonumber \\
    & \stackrel{(a)}{=}  \tr\left(\vect{R}_{{\rm g}^{\prime}}\right) \tr\left(\vect{\Phi}\ddot{\vect{U}}_{\rm B,1}\ddot{\vect{D}}_{\rm B,1}\ddot{\vect{U}}^{\Htran}_{\rm B,1}\vect{\Phi}^{\Htran}\right)
\end{align}
where $(a)$ follows from $\vect{R}_{{\rm g}^{\prime}}=\vect{U}_{{\rm g}^{\prime},1}\vect{D}_{{\rm g}^{\prime},1}\vect{U}_{{\rm g}^{\prime},1}^{\Htran}$. Now, let $\vect{\Phi}=\vect{S}_{\rm \Phi}\vect{\Lambda}_{\rm \Phi}\vect{V}_{\rm \Phi}^{\Htran}$ be the singular value decomposition of $\vect{\Phi}$. Then, we can express $\tr(\vect{A})$ in \eqref{eq:traceA} as
\begin{align} \label{eq:traceA2}
    \tr(\vect{A}) & = \tr\left(\vect{R}_{{\rm g}^{\prime}}\right) \tr\left(\vect{S}_{\rm \Phi}\vect{\Lambda}_{\rm \Phi}\vect{V}_{\rm \Phi}^{\Htran}\ddot{\vect{U}}_{\rm B,1}\ddot{\vect{D}}_{\rm B,1}\ddot{\vect{U}}^{\Htran}_{\rm B,1}\vect{V}_{\rm \Phi}\vect{\Lambda}_{\rm \Phi}^{\Ttran}\vect{S}_{\rm \Phi}^{\Htran}\right) \nonumber \\
    & = \tr\left(\vect{R}_{{\rm g}^{\prime}}\right) \tr\left(\underbrace{\vect{V}_{\rm \Phi}^{\Htran}\ddot{\vect{U}}_{\rm B}\ddot{\vect{D}}_{\rm B}\ddot{\vect{U}}^{\Htran}_{\rm B}\vect{V}_{\rm \Phi}}_{\triangleq\vect{X}}\underbrace{\vect{\Lambda}_{\rm \Phi}^{\Ttran}\vect{\Lambda}_{\rm \Phi}}_{\triangleq\vect{\Sigma}}\right)
\end{align}
where we have used the cyclic shift property of trace, the fact that the matrix $\vect{S}_{\rm \Phi}$ is unitary, and also that $\ddot{\vect{U}}_{\rm B,1}\ddot{\vect{D}}_{\rm B,1}\ddot{\vect{U}}^{\Htran}_{\rm B,1}=\ddot{\vect{U}}_{\rm B}\ddot{\vect{D}}_{\rm B}\ddot{\vect{U}}^{\Htran}_{\rm B}$. We see that the newly defined matrix inside the trace, i.e., $\vect{\Sigma}=\vect{\Lambda}_{\rm \Phi}^{\Ttran}\vect{\Lambda}_{\rm \Phi}$, is diagonal. In addition, the first matrix $\vect{X}$ satisfies $\tr(\vect{X}^{\Htran}\vect{X})=\tr(\ddot{\vect{D}}_{\rm B}^2)$, which is constant for the given RIS and EMI related correlation matrices $\vect{R}_{\rm h}$, $\vect{R}_{\rm g}$, and $\vect{R}_{\rm w}$. Using  von Neumann trace inequality, we conclude that $\vect{X}$ should be diagonal to maximize the objective function. As seen from the expression above, for the optimal solution (for $\vect{X}$ to be diagonal) we should have $\vect{V}_{\rm \Phi}=\ddot{\vect{U}}_{\rm B}$. By setting it this way, the relation in \eqref{eq:A} becomes
\begin{align} \label{eq:A-Phi}
      \vect{A}&=\left(\vect{I}_{M}\kron\vect{\Phi}\right) \left(\vect{U}_{{\rm g}^{\prime}}\kron \ddot{\vect{U}}_{\rm B}\right)\nonumber\\
      &\quad\times\left(\vect{D}_{{\rm g}^{\prime}}\kron \ddot{\vect{D}}_{\rm B}\right)\left(\vect{U}_{{\rm g}^{\prime}}^{\Htran}\kron \ddot{\vect{U}}^{\Htran}_{\rm B}\right)\left(\vect{I}_{M}\kron\vect{\Phi}^{\Htran}\right)\nonumber\\
      &=\left(\vect{U}_{{\rm g}^{\prime}}\vect{D}_{{\rm g}^{\prime}}\vect{U}_{{\rm g}^{\prime}}^{\Htran}\right)\kron\left(\vect{S}_{\rm \Phi}\vect{\Lambda}_{\Phi}\ddot{\vect{D}}_{\rm B}\vect{\Lambda}_{\Phi}^{\Ttran}\vect{S}_{\rm \Phi}^{\Htran}\right) \nonumber \\
      & = \underbrace{\left( \vect{U}_{{\rm g}^{\prime}}\kron\vect{S}_{\rm \Phi}  \right)}_{\vect{U}_{\rm A}}\underbrace{\left( \vect{D}_{{\rm g}^{\prime}}\kron\left(\vect{\Lambda}_{\rm \Phi}\ddot{\vect{D}}_{\rm B}\vect{\Lambda}_{\rm \Phi}^{\Ttran}\right)  \right)}_{\vect{D}_{\rm A}}\underbrace{\left( \vect{U}_{{\rm g}^{\prime}}^{\Htran}\kron\vect{S}_{\rm \Phi}^{\Htran}  \right)}_{\vect{U}_{\rm A}^{\Htran}}.
\end{align}
The above equation states that we can construct a valid eigendecomposition of $\vect{A}$ with the selection $\vect{V}_{\rm \Phi}=\ddot{\vect{U}}_{\rm B}$. Moreover, as a design parameter, $\vect{S}_{\rm \Phi}$  does not affect the value of the optimization objective in \eqref{eq:trace1}. Hence, it can be set as an arbitrary unitary matrix. 

Before proceeding further with the above equation, we note that $\vect{G}$ in \eqref{eq:trace1} can be expressed from \eqref{eq:Rx-lemma} as follows:
\begin{align} \label{eq:G}
    \vect{G}=\vect{F}\vect{F}^{\Htran}\vect{D}_{\rm B,1}\vect{F}\vect{F}^{\Htran}=\vect{D}_{\rm B,1}^{-\frac{1}{2}}\vect{U}_{\rm B,1}^{\Htran}\vect{R}_{\rm x}^{2}\vect{U}_{\rm B,1}\vect{D}_{\rm B,1}^{-\frac{1}{2}}.
\end{align}
From \eqref{eq:G2app} in Appendix~\ref{appendix-equation-steps2}, we can re-write $\vect{G}$ in \eqref{eq:G} as 
\begin{align}
    \label{eq:G2}
    \vect{G}&=\vect{D}_{{\rm g}^{\prime},1}\kron \underbrace{\left( \ddot{\vect{D}}_{\rm B,1}^{-\frac{1}{2}}    \ddot{\vect{U}}_{\rm B,1}^{\Htran}  \left(\vect{R}_{\rm h}\odot\vect{R}_{\rm g}\right)^2  \ddot{\vect{U}}_{\rm B,1} \ddot{\vect{D}}_{\rm B,1}^{-\frac{1}{2}}\right)}_{\triangleq \ddot{\vect{G}}\in \mathbb{C}^{\ddot{r}_{\rm B} \times \ddot{r}_{\rm B}}}.
\end{align}

Using the Kronecker product representation of $\vect{D}_{\rm A}$ in \eqref{eq:A-Phi} and $\vect{G}$ in \eqref{eq:G2}, we can re-express the objective function in \eqref{eq:trace1} as follows:
\begin{align}
  \underset{\sum_{i=1}^{\tau_p} \lambda_{{\rm \Phi}, i}^2\leq N\tau_p}{\mathacr{maximize}} \ \  \sum_{i=1}^{\ddot{r}_{\rm B}}\sum_{m=1}^M\frac{d_{{\rm g}^{\prime},m}^2\ddot{g}_{i,i}\ddot{d}_{{\rm B},i}\lambda_{{\rm \Phi},i}^2}{d_{{\rm g}^{\prime},m}\ddot{d}_{{\rm B},i}\lambda_{{\rm \Phi},i}^2+\frac{1}{\rm{SNR}}}
\end{align}
where $\ddot{g}_{i,i}$ is the $(i,i)$th entry of the matrix $\ddot{\vect{G}}$ in \eqref{eq:G2}.
The above optimization problem is equivalent to
\begin{align}
  \underset{\sum_{i=1}^{\tau_p} \lambda_{{\rm \Phi}, i}^2\leq N\tau_p}{\mathacr{minimize}} \ \  \sum_{i=1}^{\ddot{r}_{\rm B}}\sum_{m=1}^M\frac{d_{{\rm g}^{\prime},m}\ddot{g}_{i,i}}{d_{{\rm g}^{\prime},m}\ddot{d}_{{\rm B},i}\lambda_{{\rm \Phi},i}^2+\frac{1}{\rm{SNR}}}.
\end{align}

Based on the analysis outlined above, we summarize our novel result in the following lemma.

\begin{lemma} \label{lemma2b}
 Let the eigendecomposition of  $\vect{R}_{{\rm g}^{\prime}}$ and   $\ddot{\vect{B}} = \left(\vect{R}_{\rm h}\odot\vect{R}_{\rm g}\right)+\frac{1}{\rm{SIR}}\left(\vect{R}_{\rm w}\odot\vect{R}_{\rm g}\right)$  be given by $\vect{R}_{{\rm g}^{\prime}}=\vect{U}_{{\rm g}^{\prime}}\vect{D}_{{\rm g}^{\prime}}\vect{U}_{{\rm g}^{\prime}}^{\Htran}$ and $\ddot{\vect{B}}=\ddot{\vect{U}}_{\rm B}\ddot{\vect{D}}_{\rm B}\ddot{\vect{U}}_{\rm B}^{\Htran}=\ddot{\vect{U}}_{\rm B,1}\ddot{\vect{D}}_{\rm B,1}\ddot{\vect{U}}_{\rm B,1}^{\Htran}$, respectively. $\vect{D}_{{\rm g}^{\prime}}\in\mathbb{C}^{M \times M}$ and  $\ddot{\vect{D}}_{\rm B}\in\mathbb{C}^{N \times N}$  are the diagonal matrices with the real non-negative eigenvalues appearing on the diagonals in descending order, which are $d_{{\rm g}^{\prime}, 1}\geq \ldots \geq d_{{\rm g}^{\prime},M}\geq 0$ and $\ddot{d}_{{\rm B}, 1}\geq \ldots \geq \ddot{d}_{{\rm hg},N}\geq 0$, respectively. The reduced-dimension diagonal matrix $\ddot{\vect{D}}_{\rm B,1}\in \mathbb{C}^{\ddot{r}_{B}\times \ddot{r}_{B} }$ consists of the positive eigenvalues. Then the optimal $\vect{\Phi}$ has the singular value decomposition $\vect{\Phi}=\vect{S}_{\rm \Phi}\vect{\Lambda}_{\rm \Phi}\vect{V}_{\rm \Phi}^{\Htran}$ where $\vect{S}_{\rm \Phi}\in \mathbb{C}^{\tau_p\times\tau_p}$ is an arbitrary unitary matrix, $\vect{V}_{\rm \Phi}=\ddot{\vect{U}}_{\rm B}$, and the diagonal entries of $\vect{\Lambda}_{\rm \Phi}$, which are $\lambda_{{\rm \Phi},i}$, given as the solution to the problem
 \begin{align}
  \underset{\sum_{i=1}^{\tau_p} \lambda_{{\rm \Phi}, i}^2\leq N\tau_p}{\mathacr{minimize}} \ \  \sum_{i=1}^{\ddot{r}_{\rm B}}\sum_{m=1}^M\frac{d_{{\rm g}^{\prime},m}\ddot{g}_{i,i}}{d_{{\rm g}^{\prime},m}\ddot{d}_{{\rm B},i}\lambda_{{\rm \Phi},i}^2+\frac{1}{\rm{SNR}}} \label{eq:objective4}
 \end{align}
 where $\ddot{g}_{i,i}$ is the $(i,i)$th diagonal entry of the matrix $\ddot{\vect{G}}=\ddot{\vect{D}}_{\rm B,1}^{-\frac{1}{2}}    \ddot{\vect{U}}_{\rm B,1}^{\Htran}  \left(\vect{R}_{\rm h}\odot\vect{R}_{\rm g}\right)^2  \ddot{\vect{U}}_{\rm B,1} \ddot{\vect{D}}_{\rm B,1}^{-\frac{1}{2}}\in \mathbb{C}^{\ddot{r}_{\rm B} \times \ddot{r}_{\rm B}}$.
\end{lemma}
The optimization problem in \eqref{eq:objective4} for the case with EMI has a similar structure to that without EMI in \eqref{eq:objective3}, although the eigenvalues and eigenvectors are different. Hence, one can still use convex programming solvers to obtain the solution, or the water-filling algorithm if $M=1$.

Let $\vect{\Phi}^{\star}$ denote the optimal solution to \eqref{eq:lmmse-opt} from Lemma~\ref{lemma2b}. We propose to obtain the RIS phase-shift matrix with unit-modulus entries as $\vect{\Phi}=e^{\imagunit\angle{ \vect{\Phi}^{\star}}}$.

\begin{remark}Although the phase-shift projection operation results in some loss of optimality, as shown in the numerical results, it remains a low-complexity method. Alternatively, one could consider developing a majorization-minimization (MM) based algorithm using the technique proposed in \cite{liu2022joint}. However, this approach would result in a high-dimensional problem because the number of optimization variables is the product of the number of RIS elements and the pilot length. Additionally, the updates would require computing Kronecker products of large matrices. Therefore, we opt for the simple projection method in this paper. Later, we will devise an MM algorithm with closed-form updates for the optimization of the pilot matrix for the RS-LS estimator. This approach would be feasible because it involves smaller-sized matrix operations for the RS-LS estimator.
 \end{remark}
 
 Now, let us verify that the optimization problem in Lemma~\ref{lemma2b} reduces to the one in Lemma~\ref{lemma2} when there is no EMI, i.e., $\frac{1}{\rm{SIR}}=0$. In this case, the matrix $\ddot{\vect{B}}$ in Lemma~\ref{lemma2b} becomes $\ddot{\vect{B}}=\left(\vect{R}_{\rm h}\odot\vect{R}_{\rm g}\right)$ and its eigenvalue decomposition is $\ddot{\vect{B}}=\vect{U}_{\rm hg}\vect{D}_{\rm hg}\vect{U}_{\rm hg}^{\Htran}$ using the notation introduced in Lemma~\ref{lemma2}. This means that $\ddot{d}_{{\rm B},i}= d_{{\rm hg},i}$. The matrix $\ddot{\vect{G}}$ in Lemma~\ref{lemma2b} for $\frac{1}{\rm{SIR}}=0$ becomes $\ddot{\vect{G}}=\ddot{\vect{D}}_{\rm B,1}^{-\frac{1}{2}}    \ddot{\vect{U}}_{\rm B,1}^{\Htran}  \left(\vect{R}_{\rm h}\odot\vect{R}_{\rm g}\right)^2  \ddot{\vect{U}}_{\rm B,1} \ddot{\vect{D}}_{\rm B,1}^{-\frac{1}{2}}=\vect{D}_{\rm hg}^{-\frac{1}{2}}    \vect{U}_{\rm hg}^{\Htran}\vect{U}_{\rm hg}\vect{D}_{\rm hg}^2\vect{U}_{\rm hg}^{\Htran} \vect{U}_{\rm hg} \vect{D}_{\rm hg}^{-\frac{1}{2}}=\vect{D}_{\rm hg}$, which results in $\ddot{g}_{i,i}=d_{{\rm hg},i}$. When we insert $\ddot{d}_{{\rm B},i}= d_{{\rm hg},i}$ and  $\ddot{g}_{i,i}=d_{{\rm hg},i}$ into the objective function in \eqref{eq:objective4}, we obtain the same optimization problem as in \eqref{eq:objective3} for the case without EMI.

A closed-form optimal solution can be obtained for the special case of a single-antenna BS, $M=1$, as follows.

\begin{lemma}
When there is a single antenna at the BS, i.e., $M=1$, the conditions for the  optimal singular values are 
\begin{align} \label{eq:special-single3}
    \lambda_{{\rm \Phi},i} = \sqrt{\max\left(0,\frac{\sqrt{\ddot{g}_{i,i}}}{\sqrt{\ddot{d}_{{\rm B},i}}\sqrt{\mu}}-\frac{{\rm SNR}^{-1}}{\ddot{d}_{{\rm B},i}}\right)}, \quad i=1,\ldots,\tau_p
 \end{align}
 where the Lagrange multiplier $\mu>0$ satisfies the constraint
 \begin{align} \label{eq:special-single4}
     \sum_{i=1}^{\tau_p}\max\left(0,\frac{\sqrt{\ddot{g}_{i,i}}}{\sqrt{\ddot{d}_{{\rm B},i}}\sqrt{\mu}}-\frac{{\rm SNR}^{-1}}{\ddot{d}_{{\rm B},i}}\right)=N\tau_p.
 \end{align}

 \end{lemma}
 \begin{IEEEproof}
The proof follows from deriving Karush-Kuhn-Tucker conditions for the problem in \eqref{eq:objective4} with $M=1$.
 \end{IEEEproof}
 
For the case of non-zero EMI, the inter-relation between the eigenstructures of the matrices $\ddot{\vect{B}} = \left(\vect{R}_{\rm h}\odot\vect{R}_{\rm g}\right)+\frac{1}{\rm{SIR}}\left(\vect{R}_{\rm w}\odot\vect{R}_{\rm g}\right)$ and $\ddot{\vect{G}}=\ddot{\vect{D}}_{\rm B,1}^{-\frac{1}{2}}    \ddot{\vect{U}}_{\rm B,1}^{\Htran}  \left(\vect{R}_{\rm h}\odot\vect{R}_{\rm g}\right)^2  \ddot{\vect{U}}_{\rm B,1} \ddot{\vect{D}}_{\rm B,1}^{-\frac{1}{2}}$ shape the optimal phase-shift matrix. When the spatial correlation matrix of the EMI has a similar eigenspace as $\vect{R}_{\rm h}\odot\vect{R}_{\rm g}$, we expect that $\ddot{g}_{i,i}$ is close to the eigenvalue $\ddot{d}_{{\rm B},i}$, and we obtain a similar water-filling structure to the case without EMI as seen from \eqref{eq:special-single4}. For the general case, a more complex dependence exists based on the specific values of $\ddot{g}_{i,i}$ and $\ddot{d}_{{\rm B},i}$. More power is allocated to the right singular vectors of $\vect{\Phi}$ corresponding to large  $\ddot{g}_{i,i}\ddot{d}_{{\rm B},i}$ from the water-filling operation. The resulting power value is further scaled by $\sqrt{\ddot{g}_{i,i}/\ddot{d}_{{\rm B},i}}$ to give more emphasis to the desired signal space.

 \section{Optimizing the RIS Configuration for RS-LS Estimation} \label{sec:optimization-RSLS}

We will now show how to optimize the RIS matrix for the case where the RS-LS estimator is used.
Similar to the previous section, we will first consider minimizing the MSE for the case without EMI (i.e., $\frac{1}{\rm{SIR}}=0$) and then consider the general case with EMI.

\subsection{Optimal Structure  for the RS-LS Estimator Without EMI}

Relaxing the unit modulus constraints as before, our aim is to find  $\vect{\Phi}$ that minimizes ${\rm MSE}_{\rm RS-LS}^{\rm conserv}$ in \eqref{eq:mse-rsls} with $\frac{1}{\rm{SIR}}=0$. The optimization problem is
\begin{align}
& \underset{\tr\left(\vect{\Phi}^{\Htran}\vect{\Phi}\right)\leq N\tau_p}{\mathacr{minimize}} \ \ \tr\left(\left(\overline{\vect{U}}_{\rm x,1}^{\Htran}\vect{\Phi}_M^{\Htran}\vect{\Phi}_M\overline{\vect{U}}_{\rm x,1}\right)^{-1}\right).  \label{eq:rsls-opt}
\end{align}
Let the Kronecker decomposition $\overline{\vect{U}}_{\rm x,1}=\overline{\vect{U}}_{{\rm g}^{\prime},1}\kron\overline{\vect{U}}_{{\rm hg},1}$  be constructed based on the eigenspaces of the spatial correlation matrices of the BS (i.e., $\overline{\vect{R}}_{{\rm g}^{\prime}}$) and RIS (i.e., $\overline{\vect{R}}_{\rm h}\odot \overline{\vect{R}}_{\rm g}$) selected in the conservative RS-LS estimation. The objective function in \eqref{eq:rsls-opt} can be written as in \eqref{eq:rsls-opt2} at the top of the next page,
\begin{figure*}
\begin{align}
\tr\Bigg(\bigg(\left(\overline{\vect{U}}_{{\rm g}^{\prime},1}^{\Htran}\kron\overline{\vect{U}}_{{\rm hg},1}^{\Htran}\right)\left(\vect{I}_M \kron \vect{\Phi}^{\Htran}\right)\left(\vect{I}_M \kron \vect{\Phi}\right)\left(\overline{\vect{U}}_{{\rm g}^{\prime},1}\kron\overline{\vect{U}}_{{\rm hg},1}\right)\bigg)^{-1}\Bigg)& = \tr\left(\left(\vect{I}_{\overline{r}_{{\rm g}^{\prime}}}\kron \left(\overline{\vect{U}}_{{\rm hg},1}^{\Htran}\vect{\Phi}^{\Htran}\vect{\Phi}\overline{\vect{U}}_{{\rm hg},1}\right)\right)^{-1}\right) \nonumber\\
&=\overline{r}_{{\rm g}^{\prime}}\cdot\tr\left(\left(\overline{\vect{U}}_{{\rm hg},1}^{\Htran}\vect{\Phi}^{\Htran}\vect{\Phi}\overline{\vect{U}}_{{\rm hg},1}\right)^{-1}\right).\label{eq:rsls-opt2}
\end{align}
\vspace{-6mm}
\hrulefill
\end{figure*}
where we have used the distributive properties of the Kronecker product and implicitly assumed that $\vect{A}\triangleq\vect{\Phi}\overline{\vect{U}}_{{\rm hg},1}\in \mathbb{C}^{\tau_p \times \overline{r}_{\rm hg}}$ has full rank, i.e., $\rank(\vect{A})=\overline{r}_{\rm hg}$. Denoting the singular value decomposition of the matrix $\vect{A}$ as $\vect{A}=\vect{S}_{\rm A}\vect{\Lambda}_{\rm A}\vect{V}_{\rm A}^{\Htran}$ with the singular values $\lambda_{{\rm A},1}\geq \ldots \geq \lambda_{{\rm A},\overline{r}_{\rm hg}}>0$. Then, the objective function in \eqref{eq:rsls-opt2} can be expressed in terms of the singular values of $\vect{A}$ as
\begin{align}
  \overline{r}_{{\rm g}^{\prime}} \sum_{i=1}^{\overline{r}_{\rm hg}} \frac{1}{\lambda_{{\rm A},i}^2} \label{eq:objectiveb}
\end{align}
whose minimum over the singular values is a monotonically decreasing function of $\sum_{i=1}^{\overline{r}_{\rm hg}}\lambda_{{\rm A},i}^2=\tr(\vect{A}^{\Htran}\vect{A})$, which is equal to
\begin{align}
  &  \tr\left(\overline{\vect{U}}_{{\rm hg},1}\overline{\vect{U}}_{{\rm hg},1}^{\Htran}\vect{\Phi}^{\Htran}\vect{\Phi}\right)  \nonumber\\&=  \tr\left(\overline{\vect{U}}_{{\rm hg}}\overline{\vect{U}}_{{\rm hg}}^{\Htran}\vect{\Phi}^{\Htran}\vect{\Phi}\right)- \tr\left(\overline{\vect{U}}_{{\rm hg},1}\overline{\vect{U}}_{{\rm hg},2}^{\Htran}\vect{\Phi}^{\Htran}\vect{\Phi}\right) \nonumber\\
  &  =\tr\left(\vect{\Phi}^{\Htran}\vect{\Phi}\right)- \tr\left(\overline{\vect{U}}_{{\rm hg},2}\overline{\vect{U}}_{{\rm hg},2}^{\Htran}\vect{\Phi}^{\Htran}\vect{\Phi}\right) \leq N\tau_p
\end{align}
where $\overline{\vect{U}}_{{\rm hg},2}\in \mathbb{C}^{N \times (N-\overline{r}_{\rm hg})}$ is the matrix whose columns are the orthonormal eigenvectors of $\left(\overline{\vect{R}}_{{\rm h}}\odot \overline{\vect{R}}_{{\rm g}}\right)$ corresponding to zero eigenvalues. The above inequality is satisfied with equality when the right singular vectors of $\vect{\Phi}$ corresponding to non-zero singular values lie in the subspace spanned by $\overline{\vect{U}}_{{\rm hg},1}$, i.e., $\vect{\Phi}\overline{\vect{U}}_{{\rm hg},2}=\vect{0}_{\tau_p \times (N-\overline{r}_{\rm hg})}$. In this way, the objective value in \eqref{eq:objectiveb} is minimized. Moreover, $\lambda_{{\rm A},i}= \sqrt{N\tau_p/\overline{r}_{\rm hg}}$, for $i=1,\ldots,\overline{r}_{\rm hg}$ to minimize \eqref{eq:objectiveb}. We can construct the optimal $\vect{\Phi}$ that satisfies all these constraints as 
\begin{align} \label{eq:optimal-Phi-RSLS}
    \vect{\Phi}^{\star} = \sqrt{\frac{N\tau_p}{\overline{r}_{\rm hg}}}\vect{S}_{{\rm \Phi},1}\overline{\vect{U}}_{{\rm hg},1}^{\Htran}
\end{align}
where $\vect{S}_{{\rm \Phi},1}\in \mathbb{C}^{\tau_p\times \overline{r}_{\rm RIS}}$ is an arbitrary matrix with orthonormal columns. Lastly, the phase-shift matrix is obtained as $\vect{\Phi}=e^{\imagunit\angle{ \vect{\Phi}^{\star}}}$.

\subsection{Optimal Structure  for the RS-LS Estimator with EMI}
 
 Now, we consider the case $\frac{1}{\rm{SIR}}>0$ and we should consider both terms in \eqref{eq:mse-rsls} to minimize ${\rm MSE}_{\rm RS-LS}^{\rm conserv}$, which  can be expressed as in \eqref{eq:mse-rsls2} at the top of the next page,
 \begin{figure*}
\begin{align} \label{eq:mse-rsls2}
    {\rm MSE}_{\rm RS-LS}^{\rm conserv}&=\frac{\overline{r}_{{\rm g}^{\prime}}\cdot\tr\left(\left(\overline{\vect{U}}_{{\rm hg},1}^{\Htran}\vect{\Phi}^{\Htran}\vect{\Phi}\overline{\vect{U}}_{{\rm hg},1}\right)^{-1}\right)}{{\rm SNR}} \nonumber\\
    &\quad+\frac{\tr\left(\left(\overline{\vect{U}}_{\rm x,1}^{\Htran}\vect{\Phi}_M^{\Htran}\vect{\Phi}_M\overline{\vect{U}}_{\rm x,1}\right)^{-1}\overline{\vect{U}}_{\rm x,1}^{\Htran}\vect{\Phi}_M^{\Htran}\vect{\Phi}_M\left(\vect{R}_{{\rm g}^{\prime}} \kron \left(\vect{R}_{\rm w}\odot\vect{R}_{\rm g}\right)\right)\vect{\Phi}_M^{\Htran}\vect{\Phi}_M\overline{\vect{U}}_{\rm x,1}\left(\overline{\vect{U}}_{\rm x,1}^{\Htran}\vect{\Phi}_M^{\Htran}\vect{\Phi}_M\overline{\vect{U}}_{\rm x,1}\right)^{-1}\right)}{{\rm SIR}}.
\end{align}
\vspace{-6mm}
\hrulefill
\end{figure*}
where we have replaced the first term by the simplified expression in \eqref{eq:rsls-opt2}. Using the Kronecker product representation of the matrices in the second term, the MSE becomes as in \eqref{eq:mse-rsls3} at the top of the next page.
\begin{figure*}
\begin{align} \label{eq:mse-rsls3}
    &{\rm MSE}_{\rm RS-LS}^{\rm conserv}=\frac{\overline{r}_{{\rm g}^{\prime}}\cdot\tr\left(\left(\overline{\vect{U}}_{{\rm hg},1}^{\Htran}\vect{\Phi}^{\Htran}\vect{\Phi}\overline{\vect{U}}_{{\rm hg},1}\right)^{-1}\right)}{{\rm SNR}} \nonumber\\
    &+\frac{\overline{r}_{{\rm g}^{\prime}}\cdot\tr\left(\left(\overline{\vect{U}}_{\rm hg,1}^{\Htran}\vect{\Phi}^{\Htran}\vect{\Phi}\overline{\vect{U}}_{\rm hg,1}\right)^{-1}\overline{\vect{U}}_{\rm hg,1}^{\Htran}\vect{\Phi}^{\Htran}\vect{\Phi} \left(\vect{R}_{\rm w}\odot\vect{R}_{\rm g}\right)\vect{\Phi}^{\Htran}\vect{\Phi}\overline{\vect{U}}_{\rm hg,1}\left(\overline{\vect{U}}_{\rm hg,1}^{\Htran}\vect{\Phi}^{\Htran}\vect{\Phi}\overline{\vect{U}}_{\rm hg,1}\right)^{-1}\right)}{{\rm SIR}}.
\end{align}
\vspace{-6mm}
\hrulefill
\end{figure*}
Similar to the previous part, we  implicitly assume that $\vect{A}=\vect{\Phi}\overline{\vect{U}}_{{\rm hg},1}\in \mathbb{C}^{\tau_p \times \overline{r}_{\rm hg}}$ has full column rank, i.e., $\rank(\vect{A})=\overline{r}_{\rm hg}$. 
This problem is not analytically tractable, thus we will make a simplifying assumption and later show that it is valid for most of the practical scenarios. 

Inspired by the optimal result for the case without EMI in \eqref{eq:optimal-Phi-RSLS}, we assume that $\vect{\Phi}\overline{\vect{U}}_{\rm hg,2}=\vect{0}_{\tau_p\times (N-\overline{r}_{\rm hg})}$. Under this assumption, we have
\begin{align} \label{eq:PhiU1}
 \vect{\Phi}=\vect{\Phi}\left(\overline{\vect{U}}_{\rm hg,1}\overline{\vect{U}}_{\rm hg,1}^{\Htran}+\overline{\vect{U}}_{\rm hg,2}\overline{\vect{U}}_{\rm hg,2}^{\Htran}\right) = \vect{\Phi}\overline{\vect{U}}_{\rm hg,1}\overline{\vect{U}}_{\rm hg,1}^{\Htran}.
\end{align}
Inserting \eqref{eq:PhiU1}
into the MSE in \eqref{eq:mse-rsls3}, we obtain \eqref{eq:mse-rsls4} at the top of the next page,
\begin{figure*}
\begin{align} \label{eq:mse-rsls4}
    &{\rm MSE}_{\rm RS-LS}^{\rm conserv}=\frac{\overline{r}_{{\rm g}^{\prime}}\cdot\tr\left(\left(\overline{\vect{U}}_{{\rm hg},1}^{\Htran}\vect{\Phi}^{\Htran}\vect{\Phi}\overline{\vect{U}}_{{\rm hg},1}\right)^{-1}\right)}{{\rm SNR}} \nonumber\\
    &+\frac{\overline{r}_{{\rm g}^{\prime}}\cdot\tr\left(\left(\overline{\vect{U}}_{\rm hg,1}^{\Htran}\vect{\Phi}^{\Htran}\vect{\Phi}\overline{\vect{U}}_{\rm hg,1}\right)^{-2}\overline{\vect{U}}_{\rm hg,1}^{\Htran}\vect{\Phi}^{\Htran}\vect{\Phi}\overline{\vect{U}}_{\rm hg,1}\overline{\vect{U}}_{\rm hg,1}^{\Htran} \left(\vect{R}_{\rm w}\odot\vect{R}_{\rm g}\right)\overline{\vect{U}}_{\rm hg,1}\overline{\vect{U}}_{\rm hg,1}^{\Htran}\vect{\Phi}^{\Htran}\vect{\Phi}\overline{\vect{U}}_{\rm hg,1}\right)}{{\rm SIR}}.
\end{align}
\vspace{-6mm}
\hrulefill
\end{figure*}
where we have used the cyclic shift property of the trace operator. We realize that the term with the inverse operation is canceled by the other terms, thus, we end up with
\begin{align} \label{eq:mse-rsls5}
    {\rm MSE}_{\rm RS-LS}^{\rm conserv}&=\frac{\overline{r}_{{\rm g}^{\prime}}\cdot\tr\left(\left(\overline{\vect{U}}_{{\rm hg},1}^{\Htran}\vect{\Phi}^{\Htran}\vect{\Phi}\overline{\vect{U}}_{{\rm hg},1}\right)^{-1}\right)}{{\rm SNR}} \nonumber\\
    &\quad+\frac{\overline{r}_{{\rm g}^{\prime}}\cdot\tr\left(\overline{\vect{U}}_{\rm hg,1}^{\Htran} \left(\vect{R}_{\rm w}\odot\vect{R}_{\rm g}\right)\overline{\vect{U}}_{\rm hg,1}\right)}{{\rm SIR}}.
\end{align}
We see that the second term is a constant independent of the phase-shift matrix $\vect{\Phi}$. When we minimize the MSE, we obtain the same problem as in the previous section without EMI. Hence, the optimal $\vect{\Phi}$ that minimizes the MSE in  \eqref{eq:mse-rsls5} is given by  \eqref{eq:optimal-Phi-RSLS}. Then, the optimal MSE when there is a non-zero EMI includes an additional bias compared to the non-EMI case. Note that this is valid under the additional assumption $\vect{\Phi}\overline{\vect{U}}_{\rm hg,2}=\vect{0}_{\tau_p\times (N-\overline{r}_{\rm hg})}$. We describe three possible cases for which this assumption holds.
\subsubsection{When $\tau_p=\overline{r}_{\rm hg}$} We recall that there is an implicit assumption for the RS-LS channel estimator that $\vect{A}=\vect{\Phi}\overline{\vect{U}}_{{\rm hg},1}\in \mathbb{C}^{\tau_p \times \overline{r}_{\rm hg}}$ has full column rank, i.e., $\rank(\vect{A})=\overline{r}_{\rm hg}$. This leads to the requirement for the pilot length $\tau_p\geq \overline{r}_{\rm hg}$. When the pilot length takes its minimum value $\overline{r}_{\rm hg}$, to guarantee that the square matrix $\vect{\Phi}\overline{\vect{U}}_{{\rm hg},1}\in \mathbb{C}^{\overline{r}_{\rm hg} \times \overline{r}_{\rm hg}}$ is non-singular (full rank), we should have $\vect{\Phi}\overline{\vect{U}}_{\rm hg,2}=\vect{0}_{\tau_p\times (N-\overline{r}_{\rm hg})}$. In this case, our initial assumption becomes a requirement and the optimal $\vect{\Phi}$ is given as in \eqref{eq:optimal-Phi-RSLS}.

\subsubsection{When  $\overline{\vect{U}}_{\rm hg,2}^{\Htran} \left(\vect{R}_{\rm w}\odot\vect{R}_{\rm g}\right)=\vect{0}_{(N-\overline{r}_{\rm hg}) \times N}$} This corresponds to the case where the desired signal subspace, represented by $\overline{\vect{U}}_{\rm hg,1}$, includes the subspace of the EMI. In this case, we obtain the same MSE expression as in \eqref{eq:mse-rsls4} by noting that 
\begin{align}
 \left(\vect{R}_{\rm w}\odot\vect{R}_{\rm g}\right)&= \left(\overline{\vect{U}}_{\rm hg,1}\overline{\vect{U}}_{\rm hg,1}^{\Htran}+\overline{\vect{U}}_{\rm hg,2}\overline{\vect{U}}_{\rm hg,2}^{\Htran}\right)\left(\vect{R}_{\rm w}\odot\vect{R}_{\rm g}\right)\nonumber\\
 &=\overline{\vect{U}}_{\rm hg,1}\overline{\vect{U}}_{\rm hg,1}^{\Htran}\left(\vect{R}_{\rm w}\odot\vect{R}_{\rm g}\right).
\end{align}
Hence, we end up with the same optimal phase-shift matrix in \eqref{eq:optimal-Phi-RSLS} as in the case without EMI. Indeed, this scenario always holds when we utilize  $\overline{\vect{R}}_{\rm x}=\vect{R}_{\rm BS, iso}\kron\left(\vect{R}_{\rm RIS, iso}\odot\vect{R}_{\rm RIS, iso}\right)$ for which $\overline{\vect{U}}_{\rm hg,1}$ represents the eigenspace of $\left(\vect{R}_{\rm RIS, iso}\odot\vect{R}_{\rm RIS, iso}\right)$. Since isotropic scattering covers all possible channel dimensions, it automatically results in $\overline{\vect{U}}_{\rm hg,2}^{\Htran} \left(\vect{R}_{\rm w}\odot\vect{R}_{\rm g}\right)=\vect{0}_{(N-\overline{r}_{\rm hg}) \times N}$ by Lemma~\ref{lemma:span}. 

\subsubsection{When $\overline{\vect{U}}_{\rm hg,1}^{\Htran} \left(\vect{R}_{\rm w}\odot\vect{R}_{\rm g}\right)=\vect{0}_{\overline{r}_{\rm hg} \times N}$} In this case, with the assumption  $\vect{\Phi}\overline{\vect{U}}_{\rm hg,2}=\vect{0}_{\tau_p\times (N-\overline{r}_{\rm hg})}$, the EMI-related term in \eqref{eq:mse-rsls5} becomes zero, which is the lowest attainable value for this term since the trace of a positive semi-definite matrix is always non-negative. Moreover, to minimize the first term in \eqref{eq:mse-rsls5}, it is also required to have $\vect{\Phi}\overline{\vect{U}}_{\rm hg,2}=\vect{0}_{\tau_p\times (N-\overline{r}_{\rm hg})}$ as we have previously showed. Hence, the optimal solution is again obtained by \eqref{eq:optimal-Phi-RSLS} with the MSE being equal to the case without EMI. This is intuitive since $\overline{\vect{U}}_{\rm hg,1}^{\Htran} \left(\vect{R}_{\rm w}\odot\vect{R}_{\rm g}\right)=\vect{0}_{\overline{r}_{\rm hg} \times N}$ means that  $\overline{\vect{R}}_{\rm hg} \left(\vect{R}_{\rm w}\odot\vect{R}_{\rm g}\right)=\vect{0}_{N \times N}$. In this case, the subspaces of the desired signal and the EMI becomes orthogonal and the effect of EMI is fully removed with the RS-LS channel estimator. However, we should point out this is a special case.

\subsection{An MM Algorithm for Optimizing the RIS Phase-Shift Pattern}
As an alternative to projecting the optimal solution onto the unit-modulus entries, we can develop an MM algorithm using a surrogate upper bound at a given point.  Continuing with the selection of $\vect{\Phi}\overline{\vect{U}}_{{\rm hg},2}=\vect{0}_{\tau_p \times (N-\overline{r}_{\rm hg})}$, the cost function to be minimized in \eqref{eq:mse-rsls3} becomes
\begin{align}
\tr\left(\left(\overline{\vect{U}}_{{\rm hg},1}^{\Htran}\vect{\Phi}^{\Htran}\vect{\Phi}\overline{\vect{U}}_{{\rm hg},1}\right)^{-1}\right)
\end{align}
where we have omitted the constant scaling factors. Using the result from \cite[Prop.~1]{he2024joint}, we can construct a surrogate upper bound to the cost function at fixed $\vect{\Phi}_0$ in \eqref{eq:rsls-opt} as shown in \eqref{eq:rsls-opt-bound} at the top of the next page. In this equation, $a_0$, $\vect{A}_0$, and $\vect{B}_0$ are given as
\begin{align}
&    a_0 = 
3\left(\tr\left(\left(\overline{\vect{U}}_{{\rm hg},1}^{\Htran}\vect{\Phi}_0^{\Htran}\vect{\Phi}_0\overline{\vect{U}}_{{\rm hg},1}\right)^{-1}\right)\right)^2 \label{a_0}\\
& \vect{A}_0 = -\vect{\Phi}_0\overline{\vect{U}}_{{\rm hg},1}\left(\overline{\vect{U}}_{{\rm hg},1}^{\Htran}\vect{\Phi}_0^{\Htran}\vect{\Phi}_0\overline{\vect{U}}_{{\rm hg},1}\right)^{-2}\overline{\vect{U}}_{{\rm hg},1}^{\Htran}\nonumber\\
&\quad \quad -a_0\vect{\Phi}_0\overline{\vect{U}}_{{\rm hg},1}\overline{\vect{U}}_{{\rm hg},1}^{\Htran} \label{A_0} \\
& \vect{B}_0 = a_0\left(\overline{\vect{U}}_{{\rm hg},1}^*\overline{\vect{U}}_{{\rm hg},1}^{\Ttran}\kron\vect{I}_{\tau_p}\right) \label{B_0}
\end{align}
where we have used the identity $\tr(\vect{A}^{\Ttran}\vect{B}\vect{C}\vect{D}^{\Ttran})=\mathrm{vec}(\vect{A})^{\Ttran}(\vect{D}\kron\vect{B})\mathrm{vec}(\vect{C})$ in the equality in \eqref{eq:rsls-opt-bound}.
\begin{figure*}
\begin{align}
  \tr\left(\left(\overline{\vect{U}}_{\rm x,1}^{\Htran}\vect{\Phi}_M^{\Htran}\vect{\Phi}_M\overline{\vect{U}}_{\rm x,1}\right)^{-1}\right)&\leq  a_0\tr\left(\overline{\vect{U}}_{{\rm hg},1}^{\Htran}\vect{\Phi}_0^{\Htran}\vect{\Phi}_0\overline{\vect{U}}_{{\rm hg},1}\right) +2\Re\left(\tr\left(\vect{A}_0^{\Htran}\vect{\Phi}\right)\right)+\mathrm{Constant}_1 \nonumber\\
& = \mathrm{vec}(\vect{\Phi})^{\Htran}\vect{B}_0\mathrm{vec}(\vect{\Phi})+2\Re\left(\mathrm{vec}^{\Htran}(\vect{A}_0)\mathrm{vec}(\vect{\Phi})\right) +\mathrm{Constant}_1\label{eq:rsls-opt-bound}
\end{align}
\hrulefill
\end{figure*}

To obtain a closed-form update in the iterations of the MM algorithm, we also utilize the surrogate upper bound in \cite[Eq. (21)]{liu2022joint} for the quadratic function of $\mathrm{vec}(\vect{\Phi})$ in \eqref{eq:rsls-opt-bound} to obtain
\begin{align}
 &   \mathrm{vec}(\vect{\Phi})^{\Htran}\vect{B}_0\mathrm{vec}(\vect{\Phi})+2\Re\left(\mathrm{vec}(\vect{A}_0)^{\Htran}\mathrm{vec}(\vect{\Phi})\right) \nonumber\\
    &\leq a_0\mathrm{vec}^{\Htran}(\vect{\Phi})\mathrm{vec}(\vect{\Phi})+2\Re\left(\vect{b}_0^{\Htran}\mathrm{vec}(\vect{\Phi})\right)+\mathrm{Constant}_2
\end{align}
where we have used the fact that the maximum eigenvalue of $\vect{B}_0$ is $a_0$, and $\vect{b}_0$ is given as
\begin{align}
    \vect{b}_0 = \vect{B}_0\mathrm{vec}(\vect{\Phi}_0)+\mathrm{vec}(\vect{A}_0)-a_0\mathrm{vec}(\vect{\Phi}_0). \label{b_0}
\end{align}
Now imposing the unit modulus constraints, the surrogate objective function becomes
\begin{align}
2\Re\left(\vect{b}_0^{\Htran}\mathrm{vec}(\vect{\Phi})\right)
\end{align}
to be minimized since $a_0\mathrm{vec}^{\Htran}(\vect{\Phi})\mathrm{vec}(\vect{\Phi})= a_0N\tau_p$. The optimal update for a certain fixed $\vect{\Phi}_0$ is obtained as
\begin{align}
    \mathrm{vec}(\vect{\Phi})\leftarrow e^{\imagunit\angle{-\vect{b}_0}}.
\end{align}
The steps of the proposed MM algorithm are outlined in Algorithm~\ref{alg1}.

\begin{algorithm}[h!] 
	\caption{MM algorithm to minimize the RS-LS MSE.} \label{alg1}
	\begin{algorithmic}[1]
		\State {\bf Initialization:} Select $\vect{\Phi}_0$ as in \eqref{eq:optimal-Phi-RSLS} and the number of iterations $I$
		\For{$i=0,\ldots,I-1$} 
		\State Compute $a_0$, $\vect{A}_0$, $\vect{B}_0$, and $\vect{b}_0$ as in \eqref{a_0}-\eqref{B_0} and \eqref{b_0}, respectively.
		\State  $\mathrm{vec}(\vect{\Phi})\leftarrow e^{\imagunit\angle{-\vect{b}_0}}$.
  \State $\vect{\Phi}_0\leftarrow \vect{\Phi}$
		\EndFor
		\State {\bf Output:} $\vect{\Phi}$
	\end{algorithmic}
\end{algorithm}

\section{Numerical Results} \label{sec:numerical}

We will now quantify the estimation accuracy obtained with the considered estimators and RIS configuration designs. We use the normalized MSE (NMSE) as the performance metric,  which is obtained by dividing the MSE by the total average channel gain $\tr(\vect{R}_{\rm x})$. Since the average power is a constant, the solutions that minimize the MSE also minimize the NMSE. The simulation parameters are listed in Table~\ref{table} unless otherwise stated, where $\lambda$ is the wavelength. The bandwidth is 100\,kHz and the noise spectral density is $-174$\,dBm/Hz, resulting in a noise variance of $\sigma_{\rm n}^2=-124$\,dBm. The distance between the BS and RIS is 10\,m, while the distance between the RIS and the UE is 40\,m. The path loss at the reference distance of $1$\,m is given as $-35$\,dB, with a path loss exponent of $3$ for all the channels. The effect of the RIS area on path loss is accounted for by multiplying each path loss value by $A/(\lambda^2/(4\pi))$, where $A$ is the area of an RIS element and $\lambda^2/(4\pi)$ is the effective area of an isotropic antenna. The UE transmit power is $\rho=200$\,mW.  The spatial correlation matrices are generated according to the clustered scattering model in \cite[Lem.~2]{demir2022channel}  with one cluster and $\sigma_{\varphi}=\sigma_{\delta}=\pi/36$ azimuth/elevation angular standard deviations. Note that we neglect mutual coupling, whose analysis for less than $\lambda/2$ spacing antennas and RIS elements is left for future work. The SIR values are selected in the range of $[-5,15]$\,dB, consistent with the values chosen in \cite{Torres2021}. Moreover, $\tau_p$
  is at most 256 and is often much shorter during the simulations, making it feasible for relatively mobile scenarios where the number of channel uses in a coherence block is around 200–300.

\begin{table}[t!]
\vspace{-2mm}

\caption{Simulation parameters} \label{table}
\vspace{-2mm}
\begin{tabular}{c||c}
{\bf Parameter} & {\bf Value} \\ \hline
The number of BS antennas & $M=M_{\rm H}\cdot M_{\rm V}=4\cdot 4=16$ \\ \hline
  Antenna spacing at the BS    & $\lambda/4$       \\ \hline
  The number of RIS elements
     & $N=N_{\rm H}\cdot N_{\rm V}=16\cdot 16=256$   \\ \hline
      Antenna spacing at the RIS & $\lambda/8$ \\ \hline
      Nominal angles for $\vect{R}_{\rm h}$ & $\varphi=\pi/4$, $\theta=0$ \\ \hline
       Nominal angles for $\vect{R}_{\rm g^{\prime}}$ & $\varphi=\pi/4$, $\theta=0$ \\ \hline
        Nominal  angles for $\vect{R}_{\rm g}$ & $\varphi=-\pi/4$, $\theta=-\pi/6$ \\ \hline
       Nominal  angles for $\vect{R}_{\rm w}$ & $\varphi=0,\pi/5$, $\theta=-\pi/12$ \\ \hline  
\end{tabular}
\vspace{-4mm}
\end{table}

\subsection{LMMSE channel estimation}

We begin by considering LMMSE estimation along with various RIS configuration matrices. A lower bound on the NMSE in the presence of EMI is obtained by the ``LMMSE-bound'', which uses the optimal solution $\vect{\Phi}^{\star}$ from Lemma~\ref{lemma2b} in \eqref{eq:LMMSE-estimate}. This represents the lowest attainable NMSE when the cascaded channel $\vect{x}$ follows a complex Gaussian distribution, as the LMMSE estimator becomes equivalent to the MMSE estimator in this case. If we project the RIS phase-shift matrix to have unit-modulus entries, i.e., $\vect{\Phi}=e^{\imagunit\angle{ \vect{\Phi}^{\star}}}$, we obtain the proposed ``LMMSE-optimized'' scheme. We compare this proposed scheme with the benchmark ``LMMSE-optimized-noEMI'' obtained by Lemma~\ref{lemma2} (with unit modulus entries) by neglecting the EMI in the RIS phase-shift design. In addition, we consider the LMMSE estimator with random phase-shifts (LMMSE-random) as a benchmark. In the ``LMMSE-random'' case, the unit modulus entries of the RIS phase-shift matrix have angles that are independently drawn from a uniform distribution on $[0,2\pi)$.

\begin{figure}[t!]
	\hspace{-1cm}
		\begin{center}
			\includegraphics[trim={2mm 0mm 12mm 6mm},clip,width=3.2in]{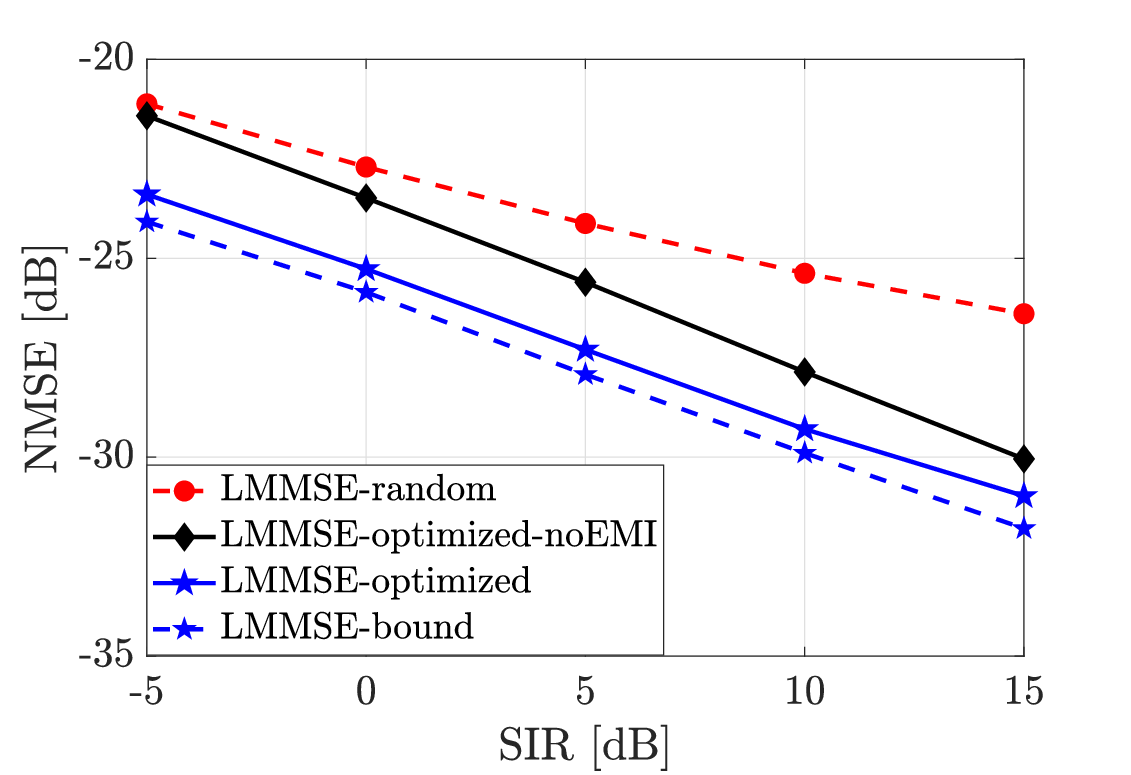}
			\vspace{-0.4cm}
			\caption{NMSE versus SIR for LMMSE estimator with $\tau_p=N/2=128$.} \label{fig:fig1}
		\end{center}
	 \vspace{-0.4cm}
\end{figure}

In Fig.~\ref{fig:fig1}, the pilot length  is  $\tau_p=N/2=128$, where we have selected it larger than the effective rank $\overline{r}_{\rm RIS}=118$ of $\overline{\vect{R}}_{\rm RIS}=\left(\vect{R}_{\rm RIS, iso}\odot\vect{R}_{\rm RIS, iso}\right)$, where the effective rank is computed as explained in Footnote~\ref{footnote1}. 
As expected, the lowest NMSE is obtained by the ``LMMSE-bound''. When we project the RIS phase-shift matrix to have unit modulus entries, LMMSE-optimized results in a small NMSE increase. However, the proposed estimator provides a substantial performance improvement compared to the random phase-shift pattern. The performance gap increases as the SIR increases reaching around 4.5\,dB reduction in NMSE when $\mathrm{SIR}=15$\,dB. Moreover, the proposed method always outperforms the scheme LMMSE-optimized-noEMI that neglects the EMI.

\begin{figure}[t!]
	\hspace{-1cm}
		\begin{center}
			\includegraphics[trim={2mm 0mm 12mm 6mm},clip,width=3.2in]{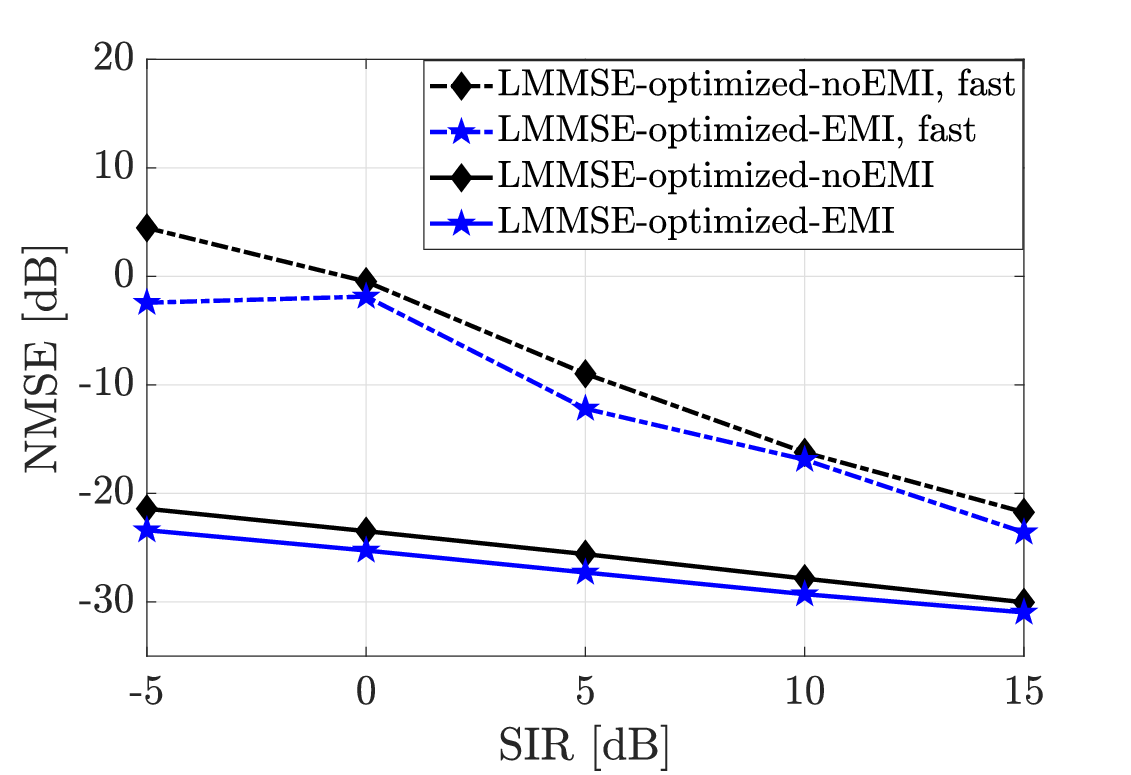}
			\vspace{-0.4cm}
			\caption{NMSE versus SIR for LMMSE estimator with fast and slowly varying EMI when the nominal azimuth angle of the EMI spatial correlation matrix is $\varphi=0$.} \label{fig:fig2}
		\end{center}
	 \vspace{-0.4cm}
\end{figure}

In Fig.~\ref{fig:fig1}, we assume that the EMI remains constant and does not change during the pilot transmission. This assumption is consistent with the analysis presented in this study. In Fig.~\ref{fig:fig2}, we replot ``LMMSE-optimized-noEMI'' and ``LMMSE-optimized-EMI'' from Fig.~\ref{fig:fig1}, which corresponds to slowly varying EMI. In addition, we include the results for rapidly varying EMI. The latter scenario is highlighted by the term ``fast'' in the figure caption and refers to the situation where the EMI undergoes a new and independent realization in each time slot during the pilot transmission. Note that the non-smooth shape for the ``LMMSE-optimized-EMI, fast'' curve is due to the RIS being optimized for slow variations. According to the figure, there is a significant performance degradation when the EMI fluctuates rapidly. On the other hand, the proposed optimization technique that exploits the EMI correlation properties achieves better performance compared to the optimized scheme that ignores the EMI. When the SIR is $5$\,dB, the optimization that exploits the EMI statistics results in an improvement of about $3$\,dB.

\begin{figure}[t!]
	\hspace{-1cm}
		\begin{center}
			\includegraphics[trim={2mm 0mm 12mm 6mm},clip,width=3.2in]{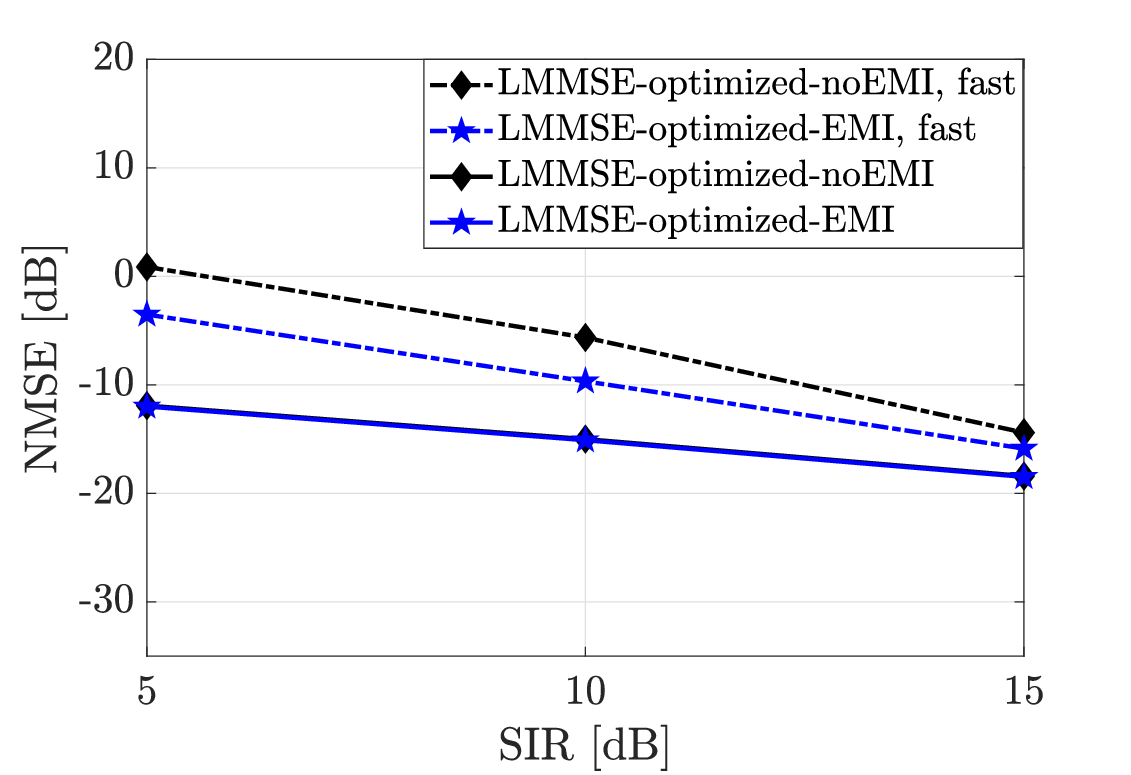}
			\vspace{-0.4cm}
			\caption{NMSE versus SIR for LMMSE estimator with fast and slowly varying EMI when the nominal azimuth angle of the EMI spatial correlation matrix is $\varphi=\pi/5$.} \label{fig:fig3}
		\end{center}
	 \vspace{-0.5cm}
\end{figure}

So far, we have assumed that the azimuth direction of the EMI (local scattering around $\varphi=0$) is significantly different from the azimuth direction of the UE channel (local scattering around $\varphi=\pi/4$). In Fig.~\ref{fig:fig3}, we examine the same simulation setup as before, but this time we modify the nominal azimuth angle of the EMI spatial correlation matrix, denoted as $\vect{R}_{\rm w}$, to $\varphi=\pi/5$. This adjustment brings the angle closer to the azimuth direction of the UE channel. Examining Fig.~\ref{fig:fig3}, it is evident that the NMSE experiences a significant increase compared to the configuration shown in Fig.~\ref{fig:fig2}. This phenomenon occurs as a direct result of the increased challenge in mitigating the EMI when the orientations of the UE and EMI channels are closer together. When the EMI remains constant during the pilot transmission, the optimized technique based on the EMI characteristics shows only a marginal improvement over the optimization scheme that ignores EMI. However, when the EMI changes rapidly, optimizing the phase-shift configuration in the RIS by considering the spatial statistics of the EMI can lead to a significant performance improvement.

\begin{figure}[t!]
	\hspace{-1cm}
		\begin{center}
			\includegraphics[trim={2mm 0mm 12mm 6mm},clip,width=3.2in]{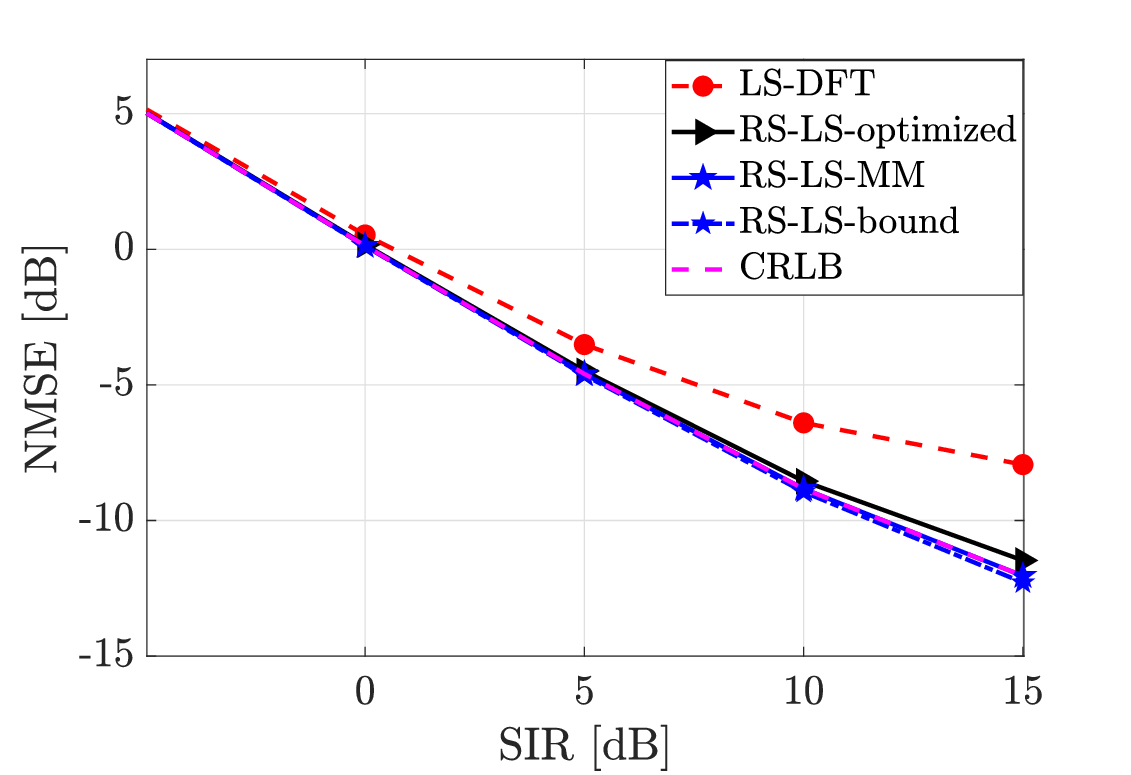}
			\vspace{-0.4cm}
			\caption{NMSE versus SIR for LS and RS-LS estimators with $\tau_p=N=256$ and when the nominal azimuth angle of the EMI spatial correlation matrix is $\varphi=\pi/5$.} \label{fig:fig4}
		\end{center}
	 \vspace{-0.5cm}
\end{figure}

\subsection{Conservative RS-LS channel estimators}

We now consider the conservative RS-LS estimator in \eqref{eq:RS-LS-estimate-approx} with the eigenspace ($\overline{\vect{U}}_{\rm x,1}$) constructed with the spatial correlation matrices for isotropic scattering, namely $\vect{R}_{\rm BS, iso}$ and $\vect{R}_{\rm RIS, iso}$ from \eqref{R-iso}, respectively. As before, ``bound'' in the legends represents the NMSE when we do not project the optimal RIS phase-shift matrix from \eqref{eq:optimal-Phi-RSLS} to have unit-modulus entries, while ``optimized'' represents the proposed scheme after the projection. RS-LS-MM denotes the optimized scheme using the MM algorithm outlined in Algorithm~\ref{alg1}.
 In addition, we consider two benchmarks: i) the conventional LS estimator with the discrete Fourier transform (DFT) matrix as the RIS phase-shift matrix; ii) the conservative RS-LS estimator with random RIS phase-shifts.  The ``DFT'' case corresponds to one of the alternative optimal phase-shift matrices when $\tau_p=N=\overline{r}_{\rm RIS}$ from \eqref{eq:optimal-Phi-RSLS}. We also plot the Cram\'{e}r-Rao lower bound (CRLB) \cite{kay_complex_crlb} using the phase-shift matrix optimized by the MM algorithm.

 In Fig.~\ref{fig:fig4}, the pilot length is set to $\tau_p=N$. Since the DFT phase-shift matrix outperforms the random phase-shift matrix for the LS estimator in this setup, we only show the results with the DFT phase-shift matrix in this figure.  As expected, the NMSE is higher than when using the LMMSE estimator (Fig.~\ref{fig:fig1}) due to the lack of user-specific spatial correlation information.  On the other hand, the conservative RS-LS method with the DFT matrix as the RIS phase-shift matrix significantly reduces the NMSE compared to the conventional LS estimator by exploiting the spatial correlation induced by the array geometries of BS and RIS. The proposed RS-LS estimator, with the phase-shift matrix optimized using the MM algorithm, closely matches the CRLB and achieves up to 4.1\,dB lower NMSE than the conventional LS estimator at high SIR levels. When the SIR is small, the choice of phase-shift matrix has little impact on the NMSE. We also observe that projecting the optimal solution onto unit-modulus entries (i.e., RS-LS-optimized) achieves performance that is sufficiently close to that of RS-LS-MM, but with significantly lower optimization complexity.

\begin{figure}[t!]
	\hspace{-1cm}
		\begin{center}
			\includegraphics[trim={2mm 0mm 12mm 6mm},clip,width=3.2in]{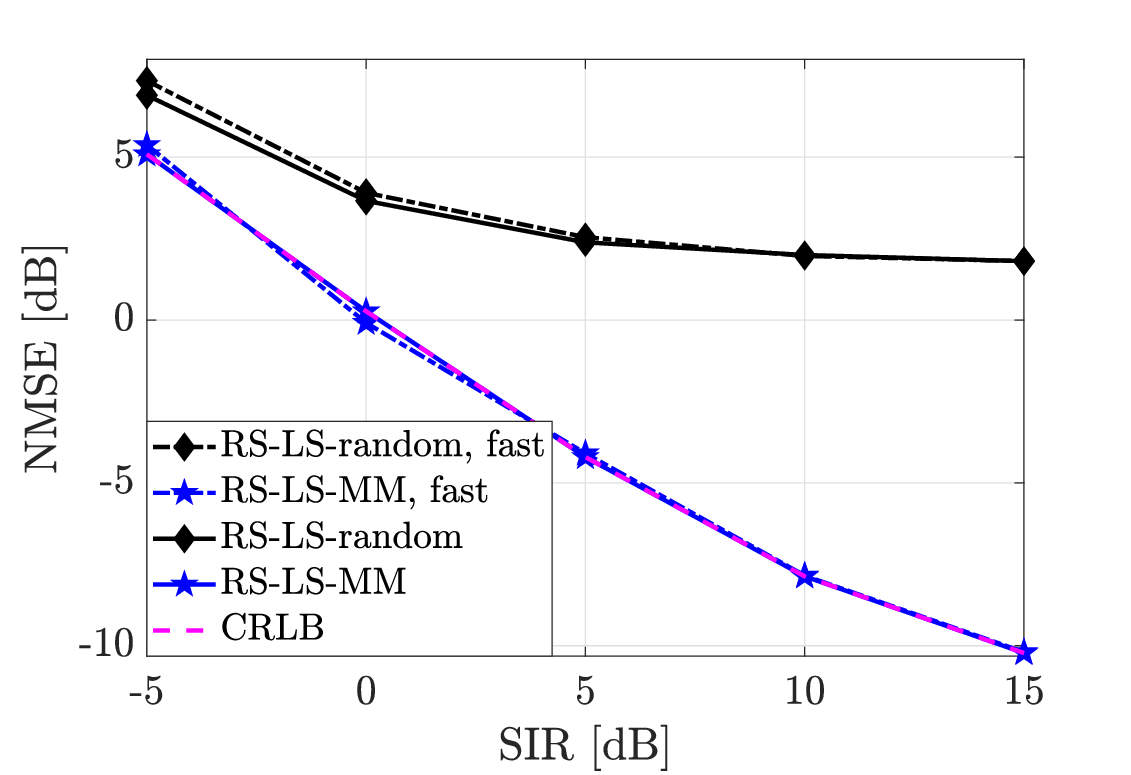}
			\vspace{-0.4cm}
			\caption{NMSE versus SIR for RS-LS estimator with  fast and slowly varying EMI when the nominal azimuth angle of the EMI spatial correlation matrix is $\varphi=\pi/5$ and $\tau_p=N/2=128$.} \label{fig:fig5}
		\end{center}
	 \vspace{-0.5cm}
\end{figure}

Fig.~\ref{fig:fig5} considers a reduced pilot length $\tau_p=N/2=128$. Due to $\tau_p<N$ and the resulting rank deficiency in \eqref{eq:RS-LS-estimate-approx}, the conventional LS estimator and the RS-LS estimator with DFT configuration cannot be used in the setup of Fig.~\ref{fig:fig5}. On the other hand, the proposed RS-LS estimator provides reasonable channel estimation accuracy when $\tau_p$ is less than half of $N$. We consider slow and fast varying EMI. For both EMI models, the gap between the randomized and optimized phase shifts is more than $11.5$\,dB at high SIR, proving the effectiveness of the proposed optimized phase-shift method. We observe once again that the performance of the proposed RS-LS-MM is very close to the CRLB computed using the same optimized phase-shift matrix.

\begin{figure}[t!]
	\hspace{-1cm}
		\begin{center}
			\includegraphics[trim={2mm 0mm 12mm 6mm},clip,width=3.2in]{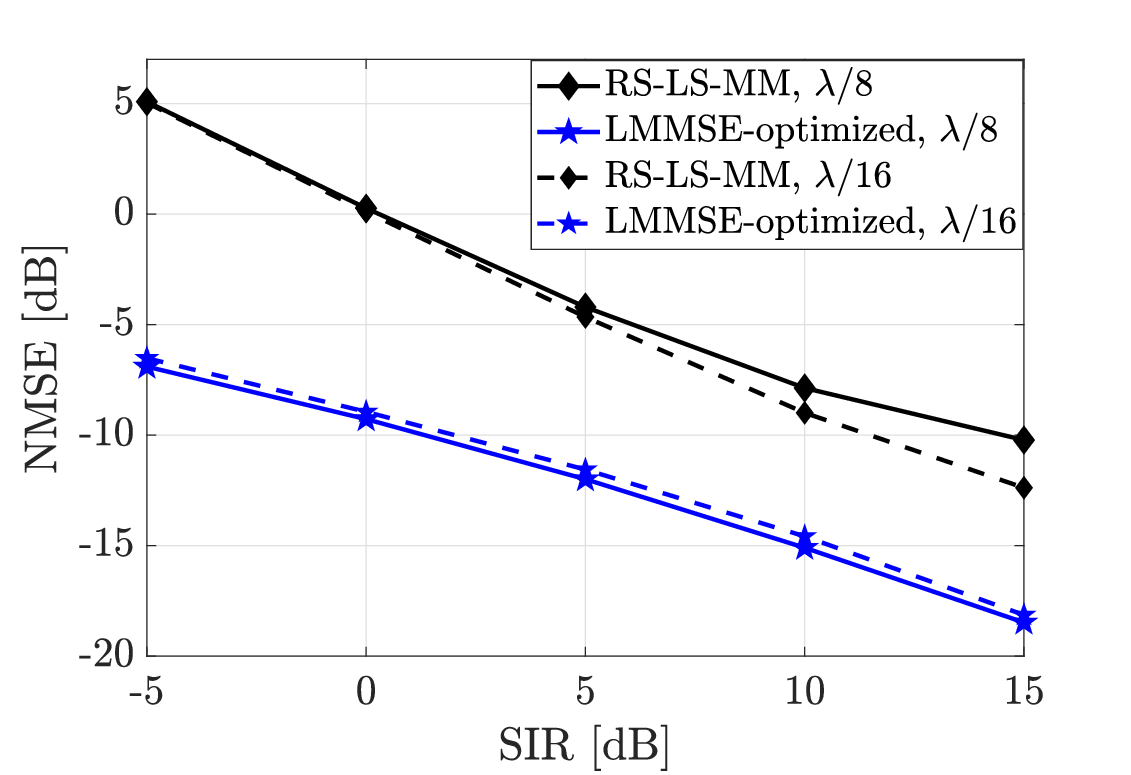}
			\vspace{-0.4cm}
			\caption{NMSE versus SIR for LMMSE and RS-LS estimators with different RIS element spacings.} \label{fig:fig6}
		\end{center}
	 \vspace{-0.5cm}
\end{figure}

\subsection{Impact of RIS Element Spacing}

In Fig.~\ref{fig:fig6}, we consider the previously selected $\lambda/8$ RIS element spacing with $\tau_p=N/2$ and additionally a denser surface with $\lambda/16$ spacing. For the denser surface case $\overline{r}_{\rm RIS}=51$ (obtained as explained in Footnote~\ref{footnote1}) and therefore we reduce the number of pilot sequences to $\tau_p=N/4=64$ for this case. Although the pilot length is reduced, a similar NMSE is obtained for both LMMSE and RS-LS estimators due to the lower rank of the spatial correlation matrices. This result is consistent with the fact that the rank of a $N \times N$ spatial correlation matrix is proportional to $d^2$ when the inter-element distance $d$ is small \cite[Prop.~2]{Bjornson2021b}. Thus, for a smaller inter-element distance (i.e., $\lambda/16$) but the same $N=256$, the RIS spatial correlation matrices have lower ranks and the channel estimation accuracy is improved by exploiting the column spaces of the RIS spatial correlation matrices more effectively. However, the reduction in the effective pilot SNR and path loss (due to smaller RIS element area) results in approximately the same NMSE performance for different antenna spacings.

\subsection{Impact of RIS Phase-Shift Quantization}
\begin{figure}[t!]
	\hspace{-1cm}
		\begin{center}
			\includegraphics[trim={2mm 0mm 12mm 6mm},clip,width=3.2in]{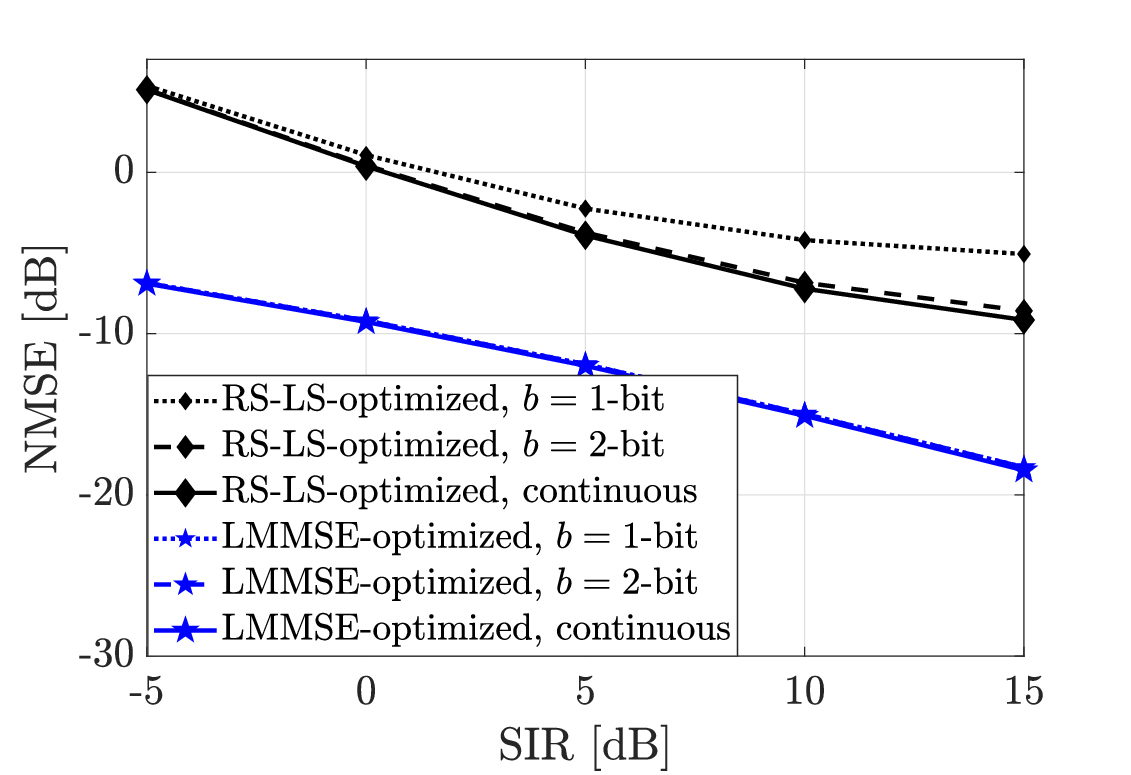}
			\vspace{-0.4cm}
			\caption{NMSE versus SIR for LMMSE and RS-LS estimators with different RIS phase-shift quantization levels.} \label{fig:fig7}
		\end{center}
	 \vspace{-0.5cm}
\end{figure}

Until now, we have assumed a continuous phase-shift for the RIS elements, which can be practically implemented using a varactor-based RIS by adjusting the bias voltage of the varactor \cite{fara2021reconfigurable,Wolff2023Continuous}. However, to analyze the impact of quantizing the RIS phase-shifts, we consider $b=1$-bit and $b=2$-bit phase-shift resolutions for the LMMSE and RS-LS estimators, as shown in Fig~\ref{fig:fig7}. We then compare the resulting NMSE with the ideal continuous phase-shifts. The quantized solutions are obtained by adjusting the angles of the elements of the optimal RIS phase-shift pattern. As illustrated in the figure, the LMMSE estimator is more robust to limited quantization bits compared to the RS-LS estimator. Nonetheless, with a $b=2$-bit resolution, it is possible to achieve NMSE performance comparable to continuous phase-shift values for both estimators.

\subsection{Impact of Number of RIS Elements and BS Antennas}

\begin{figure}[t!]
	\hspace{-1cm}
		\begin{center}
			\includegraphics[trim={2mm 0mm 12mm 6mm},clip,width=3.2in]{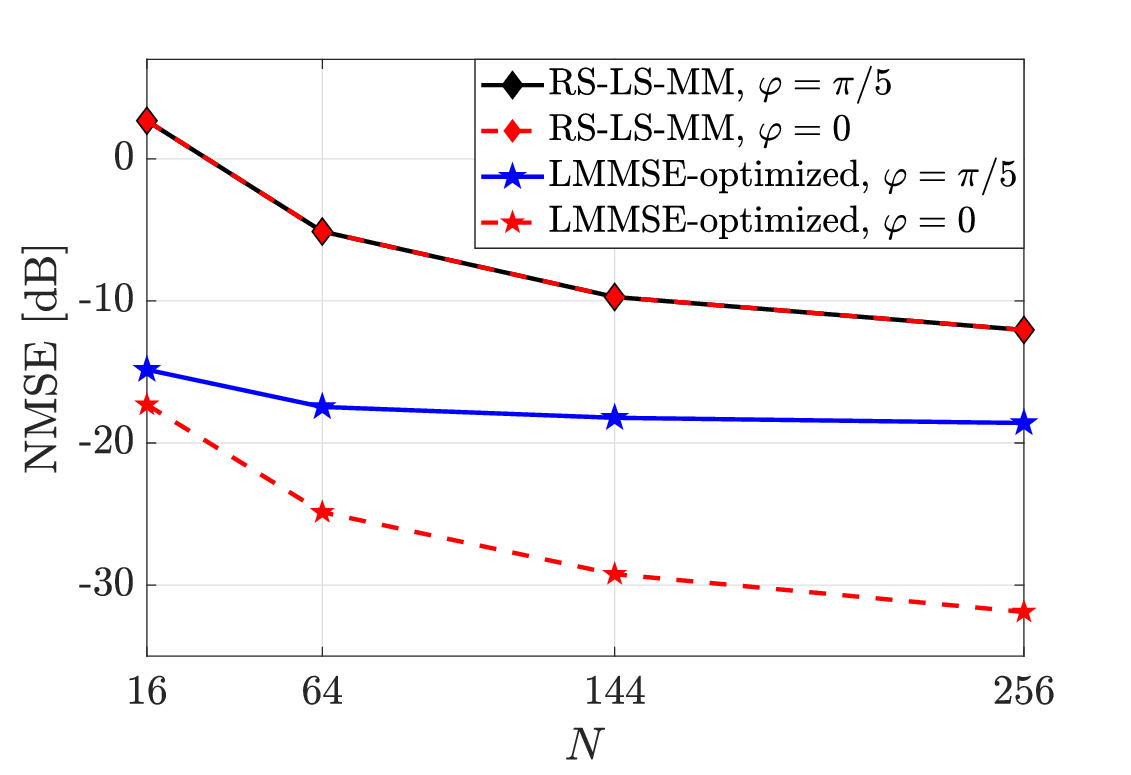}
			\vspace{-0.4cm}
			\caption{NMSE versus the number of RIS elements, $N$, for LMMSE and RS-LS estimators with different EMI angles.} \label{fig:fig8}
		\end{center}
	 \vspace{-0.5cm}
\end{figure}

In Fig.~\ref{fig:fig8}, we maintain the number of BS antennas at $M=16$ and the RIS element spacing at $\lambda/8$, while varying the number of RIS elements. We consider two angular directions for the EMI: (i) $\varphi=0$ (where the EMI is weaker) and (ii) $\varphi=\pi/5$ (where the EMI is stronger). Regardless of the EMI direction, the RS-LS estimator performs almost consistently but shows a decrease as $N$ increases. In contrast, the performance of the LMMSE estimator improves with increasing $N$, allowing for better exploitation of spatial correlation. Notably, the LMMSE estimator is more effective at rejecting EMI when the EMI direction is farther from the UE's direction.

\begin{figure}[t!]
	\hspace{-1cm}
		\begin{center}
			\includegraphics[trim={2mm 0mm 12mm 6mm},clip,width=3.2in]{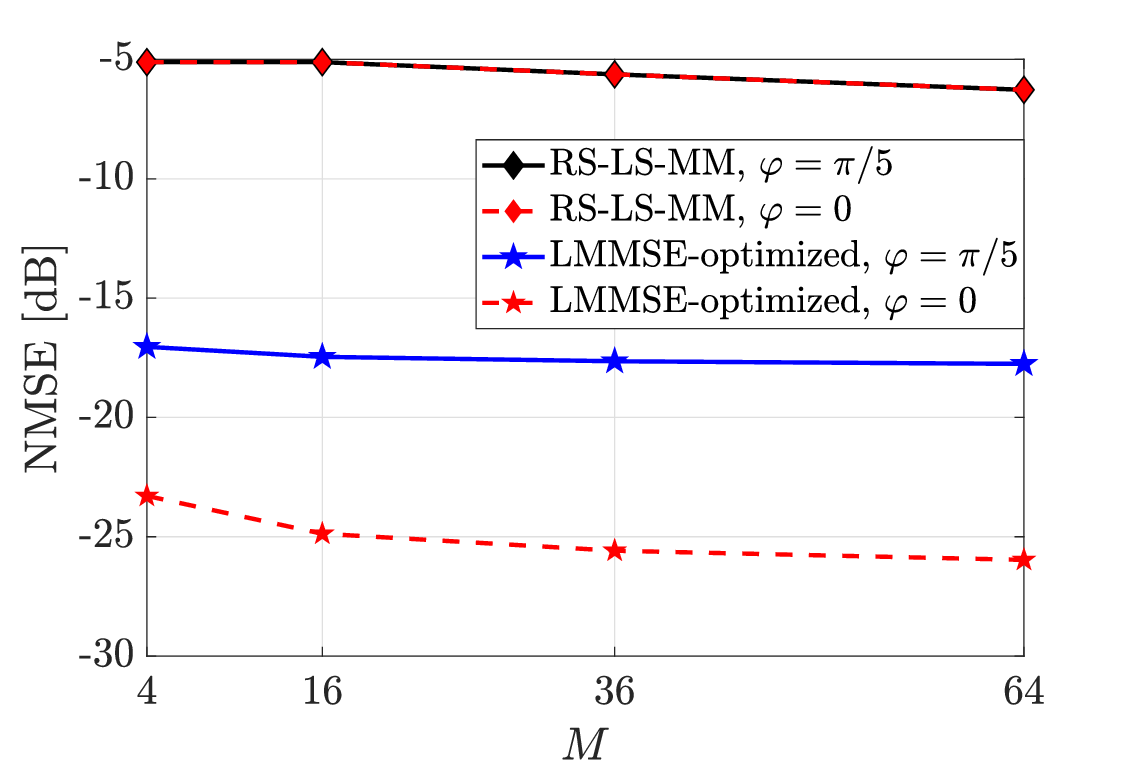}
			\vspace{-0.4cm}
			\caption{NMSE versus the number of BS antennas, $M$, for LMMSE and RS-LS estimators with different EMI angles.} \label{fig:fig9}
		\end{center}
	 \vspace{-0.5cm}
\end{figure}

In Fig.~\ref{fig:fig9}, we fix $N$ at 64 and vary $M$. Although increasing $M$ improves estimation performance, the reduction in NMSE is marginal compared to the previous figure where RIS elements were varied. This indicates that both estimators exploit the RIS correlation matrices or subspaces more effectively than the BS-side correlation or subspaces.

\section{Conclusions}\label{sec:conclusions}

A major concern in RIS-aided channel estimation is the high pilot dimensionality. We first considered the LMMSE estimator and optimized the RIS configurations by taking statistical spatial channel correlation into account, thereby allowing for jointly reducing the pilot length and improving estimation performance. The performance gain over conventional RIS phase-shift configurations is particularly large when the electromagnetic interference has a distinguishable structure. 

The downside with the LMMSE estimator is that the UE-specific statistics might be hard to acquire in practice. To address this issue, we proposed a novel RS-LS channel estimator, which exploits the array geometry and the resulting low-rank structure that any user channel will have. We optimized RIS configurations to minimize the MSE. The proposed estimator greatly outperforms the LS estimator and can be utilized with a much shorter training pilot length.  When the pilot length is small, optimizing the RIS configuration during the pilot signaling is necessary to obtain accurate channel estimates. One future direction is to consider multiple-antenna UEs by exploiting the spatial correlation and subspace characteristics on the UE array side.

\appendices

\section{Proof of Lemma~\ref{lemma:useful1}} \label{appendix-lemma-useful1}

Recall that the eigendecomposition of $\vect{R}_{\rm x}$ is given by $\vect{R}_{\rm x}=\vect{U}_{\rm x,1}\vect{D}_{\rm x,1}\vect{U}_{\rm x,1}^{\Htran}$ where $\vect{U}_{\rm x,1}\in \mathbb{C}^{MN\times r_{\rm x}}$ has the same column space as $\vect{R}_{\rm x}$. To prove the lemma, we need to show that we can express $\vect{U}_{\rm x,1}\vect{D}_{\rm x,1}^{\frac{1}{2}}$ as
\begin{align}
    \vect{U}_{\rm x,1}\vect{D}_{\rm x,1}^{\frac{1}{2}} = \vect{U}_{\rm B,1}\vect{D}_{\rm B,1}^{\frac{1}{2}}\vect{F}
\end{align}
which is equivalent to showing that the column space of $\vect{U}_{\rm x,1}$ is a subspace of the column space of $\vect{U}_{\rm B,1}$ since the diagonal matrices $\vect{D}_{\rm x,1}\in\mathbb{C}^{r_{\rm x}\times r_{\rm x}}$ and  $\vect{D}_{\rm B,1}\in\mathbb{C}^{r_{\rm B}\times r_{\rm B}}$ are non-singular. This in turn is the same as proving that the column space of $\vect{R}_{\rm x}$ is a subspace of the column space of $\vect{B}$. To show this, let $\vect{x}\in \mathbb{C}^{MN}$ be an arbitrary vector that lies in the nullspace of $\vect{B}$. Then, we have
\begin{align}
    \vect{x}^{\Htran}\vect{B}\vect{x} = 0 =  \underbrace{\vect{x}^{\Htran}\vect{R}_{\rm x}\vect{x}}_{\geq 0}+\frac{1}{\rm{SIR}}\underbrace{\vect{x}^{\Htran}\vect{R}_{\rm z}\vect{x}}_{\geq 0}
\end{align}
where each term on the right-hand side of the equation is non-negative since $\vect{R}_{\rm x}$ and $\vect{R}_{\rm z}$ are correlation matrices, and, hence positive semi-definite. To satisfy the above equation, we thus conclude that $\vect{x}^{\Htran}\vect{R}_{\rm  x}\vect{x}=0$, which in turn implies that $\vect{x}$ is also in the nullspace of $\vect{R}_{\rm x}$. Hence, no vector outside the column space of $\vect{R}_{\rm B}$ can belong to the column space of $\vect{R}_{\rm x}$, which concludes the proof.

\section{Derivation of~\eqref{eq:trace1}} 
\label{appendix-equation-steps}

Using  \eqref{eq:B}, \eqref{eq:Rx-lemma}, and \eqref{eq:A}, the objective function in \eqref{eq:lmmse-opt} can be written as in \eqref{eq:trace1app} at the top of the next page,
\begin{figure*}
    \begin{align}
&\tr\left(\!\!\vect{U}_{\rm B,1}\vect{D}_{\rm B,1}^{\frac{1}{2}}\vect{F}\vect{F}^{\Htran}\!\underbrace{\vect{D}_{\rm B,1}^{\frac{1}{2}}\vect{U}_{\rm B,1}^{\Htran}\vect{\Phi}_M^{\Htran}}_{\vect{D}_{\rm A,1}^{\frac{1}{2}}\vect{U}_{\rm A,1}^{\Htran}}\!\!\left(\underbrace{\vect{\Phi}_M\vect{U}_{\rm B,1}\vect{D}_{\rm B,1}\vect{U}_{\rm B,1}^{\Htran}\vect{\Phi}_M^{\Htran}}_{\vect{U}_{\rm A}\vect{D}_{\rm A}\vect{U}_{\rm A}^{\Htran}}+\frac{1}{\rm{SNR}}\vect{I}_{M\tau_p}\!\!\right)^{\!\!-1}\!\!\!\!\!\underbrace{\vect{\Phi}_M\vect{U}_{\rm B,1}\vect{D}_{\rm B,1}^{\frac{1}{2}}}_{\vect{U}_{\rm A,1}\vect{D}_{\rm A,1}^{\frac{1}{2}}}\!\vect{F}\vect{F}^{\Htran}\vect{D}_{\rm B,1}^{\frac{1}{2}}\vect{U}_{\rm B,1}^{\Htran}\!\right) \nonumber \\
  \stackrel{(a)}{=}& \tr\left(\underbrace{\vect{F}\vect{F}^{\Htran}\vect{D}_{\rm B,1}\vect{F}\vect{F}^{\Htran}}_{\triangleq\vect{G}\in \mathbb{C}^{r_{\rm B} \times r_{\rm B}} }\vect{D}_{\rm A,1}^{\frac{1}{2}}\vect{U}_{\rm A,1}^{\Htran}\left(\vect{U}_{\rm A}\left(\vect{D}_{\rm A}+\frac{1}{\rm{SNR}}\vect{I}_{M\tau_p}\right)\vect{U}_{\rm A}^{\Htran}\right)^{-1}\vect{U}_{\rm A,1}\vect{D}_{\rm A,1}^{\frac{1}{2}}\right)   \nonumber \\
    \stackrel{(b)}{=}& \tr\left(\vect{G}\vect{D}_{\rm A,1}^{\frac{1}{2}}\vect{U}_{\rm A,1}^{\Htran}\vect{U}_{\rm A}\left(\vect{D}_{\rm A}+\frac{1}{\rm{SNR}}\vect{I}_{M\tau_p}\right)^{-1}\vect{U}_{\rm A}^{\Htran}\vect{U}_{\rm A,1}\vect{D}_{\rm A,1}^{\frac{1}{2}}\right)   
\stackrel{(c)}{=}   \sum_{i=1}^{r_{\rm B}}\frac{g_{i,i}d_{{\rm A},i}}{d_{{\rm A},i}+\frac{1}{\rm{SNR}}}. \label{eq:trace1app}
\end{align}
\vspace{-6mm}
\hrulefill
\end{figure*}
where we have used the cyclic shift of the trace operator and recognized $\vect{A}$ from \eqref{eq:A} in $(a)$. In $(a)$, we also define the matrix $\vect{G}$. In $(b)$, we have used $\vect{U}_{\rm A}\vect{U}_{\rm A}^{\Htran}=\vect{U}_{\rm A}^{\Htran}\vect{U}_{\rm A}=\vect{I}_{M\tau_p}$. Then, noting that the matrix $\vect{D}_{\rm A,1}^{\frac{1}{2}}\vect{U}_{\rm A,1}^{\Htran}\vect{U}_{\rm A}\left(\vect{D}_{\rm A}+\frac{1}{\rm{SNR}}\vect{I}_{M\tau_p}\right)^{-1}\vect{U}_{\rm A}^{\Htran}\vect{U}_{\rm A,1}\vect{D}_{\rm A,1}^{\frac{1}{2}}$ is diagonal, we obtain the summation in $(c)$, where $g_{i,i}\geq0$ and $d_{{\rm A},i}\geq 0$ denote the $(i,i)$th entry of $\vect{G}$ and $\vect{D}_{\rm A,1}$, respectively.

\section{Appendix: Derivation of~\eqref{eq:G2}} 
\label{appendix-equation-steps2}

Recalling the Kronecker products in \eqref{eq:B-eigenvalue} and the definition of $\vect{R}_{\rm x}$ in \eqref{eq:Rx}, we can re-write $\vect{G}$ in \eqref{eq:G} as 
\begin{align}
    \label{eq:G2app}
    &\vect{G}=\left(\vect{D}_{{\rm g}^{\prime},1}^{-\frac{1}{2}}\kron \ddot{\vect{D}}_{\rm B,1}^{-\frac{1}{2}}\right)\left(\vect{U}_{{\rm g}^{\prime},1}^{\Htran}\kron \ddot{\vect{U}}_{\rm B,1}^{\Htran}\right)\nonumber\\
    &\times\left(\vect{R}_{{\rm g}^{\prime}}^2 \kron \left(\vect{R}_{\rm h}\odot\vect{R}_{\rm g}\right)^2\right)\left(\vect{U}_{{\rm g}^{\prime},1}\kron \ddot{\vect{U}}_{\rm B,1}\right)\left(\vect{D}_{{\rm g}^{\prime},1}^{-\frac{1}{2}}\kron \ddot{\vect{D}}_{\rm B,1}^{-\frac{1}{2}}\right) \nonumber\\
    & = \left(\vect{D}_{{\rm g}^{\prime},1}^{-\frac{1}{2}} \vect{U}_{{\rm g}^{\prime},1}^{\Htran} \vect{R}_{{\rm g}^{\prime}}^2 \vect{U}_{{\rm g}^{\prime},1}  \vect{D}_{{\rm g}^{\prime},1}^{-\frac{1}{2}}\right)\nonumber\\
    &\kron\underbrace{\left( \ddot{\vect{D}}_{\rm B,1}^{-\frac{1}{2}}    \ddot{\vect{U}}_{\rm B,1}^{\Htran}  \left(\vect{R}_{\rm h}\odot\vect{R}_{\rm g}\right)^2  \ddot{\vect{U}}_{\rm B,1} \ddot{\vect{D}}_{\rm B,1}^{-\frac{1}{2}}\right)}_{\triangleq \ddot{\vect{G}}\in \mathbb{C}^{\ddot{r}_{\rm B} \times \ddot{r}_{\rm B}}}  =\vect{D}_{{\rm g}^{\prime},1}\kron \ddot{\vect{G}}.
\end{align}

\bibliographystyle{IEEEtran}
\bibliography{IEEEabrv,refs}

\begin{thebibliography}{10}
\providecommand{\url}[1]{#1}
\csname url@samestyle\endcsname
\providecommand{\newblock}{\relax}
\providecommand{\bibinfo}[2]{#2}
\providecommand{\BIBentrySTDinterwordspacing}{\spaceskip=0pt\relax}
\providecommand{\BIBentryALTinterwordstretchfactor}{4}
\providecommand{\BIBentryALTinterwordspacing}{\spaceskip=\fontdimen2\font plus
\BIBentryALTinterwordstretchfactor\fontdimen3\font minus
  \fontdimen4\font\relax}
\providecommand{\BIBforeignlanguage}[2]{{%
\expandafter\ifx\csname l@#1\endcsname\relax
\typeout{** WARNING: IEEEtran.bst: No hyphenation pattern has been}%
\typeout{** loaded for the language `#1'. Using the pattern for}%
\typeout{** the default language instead.}%
\else
\language=\csname l@#1\endcsname
\fi
#2}}
\providecommand{\BIBdecl}{\relax}
\BIBdecl

\bibitem{demir2022exploiting}
{\"O}.~T. Demir, E.~Bj{\"o}rnson, and L.~Sanguinetti, ``Exploiting array
  geometry for reduced-subspace channel estimation in {RIS}-aided
  communications,'' in \emph{Proc. IEEE Sens. Array Multichannel Signal
  Process. Workshop (SAM)}, 2022, pp. 455--459.

\bibitem{Wu2019}
Q.~{Wu} and R.~{Zhang}, ``Intelligent reflecting surface enhanced wireless
  network via joint active and passive beamforming,'' \emph{IEEE Trans. Wirel.
  Commun.}, vol.~18, no.~11, pp. 5394--5409, 2019.

\bibitem{RISchannelEstimation_nested_knownBSRIS2}
Q.~{Nadeem}, H.~{Alwazani}, A.~{Kammoun}, A.~{Chaaban}, M.~{Debbah}, and
  M.~{Alouini}, ``Intelligent reflecting surface-assisted multi-user {MISO}
  communication: Channel estimation and beamforming design,'' \emph{IEEE Open
  J. Commun. Soc.}, vol.~1, pp. 661--680, 2020.

\bibitem{pei2021ris}
X.~Pei, H.~Yin, L.~Tan, L.~Cao, Z.~Li, K.~Wang, K.~Zhang, and E.~Bj{\"o}rnson,
  ``{RIS}-aided wireless communications: Prototyping, adaptive beamforming, and
  indoor/outdoor field trials,'' \emph{IEEE Trans. Commun.}, vol.~69, no.~12,
  pp. 8627--8640, 2021.

\bibitem{Bjornson2022a}
E.~Bj\"{o}rnson, H.~Wymeersch, B.~Matthiesen, P.~Popovski, L.~Sanguinetti, and
  E.~de~Carvalho, ``Reconfigurable intelligent surfaces: A signal processing
  perspective with wireless applications,'' \emph{IEEE Signal Process. Mag.},
  vol.~39, no.~2, pp. 135--158, 2022.

\bibitem{RIS_emil_magazine}
E.~Bj\"{o}rnson, {\"{O}}.~\"{O}zdogan, and E.~G. {Larsson}, ``Reconfigurable
  intelligent surfaces: Three myths and two critical questions,'' \emph{IEEE
  Commun. Mag.}, vol.~58, no.~12, pp. 90--96, 2020.

\bibitem{xu2023reconfiguring}
J.~Xu, C.~Yuen, C.~Huang, N.~Ul~Hassan, G.~C. Alexandropoulos, M.~Di~Renzo, and
  M.~Debbah, ``Reconfiguring wireless environments via intelligent surfaces for
  {6G}: Reflection, modulation, and security,'' \emph{Science China Information
  Sciences}, vol.~66, no.~3, p. 130304, 2023.

\bibitem{mishra2019channel}
D.~Mishra and H.~Johansson, ``Channel estimation and low-complexity beamforming
  design for passive intelligent surface assisted {MISO} wireless energy
  transfer,'' in \emph{ICASSP 2019-2019 IEEE International Conference on
  Acoustics, Speech and Signal Processing (ICASSP)}.\hskip 1em plus 0.5em minus
  0.4em\relax IEEE, 2019, pp. 4659--4663.

\bibitem{you2020channel}
C.~You, B.~Zheng, and R.~Zhang, ``Channel estimation and passive beamforming
  for intelligent reflecting surface: Discrete phase shift and progressive
  refinement,'' \emph{IEEE Journal on Selected Areas in Communications},
  vol.~38, no.~11, pp. 2604--2620, 2020.

\bibitem{alwazani2020intelligent}
H.~Alwazani, A.~Kammoun, A.~Chaaban, M.~Debbah, M.-S. Alouini \emph{et~al.},
  ``Intelligent reflecting surface-assisted multi-user miso communication:
  Channel estimation and beamforming design,'' \emph{IEEE Open Journal of the
  Communications Society}, vol.~1, pp. 661--680, 2020.

\bibitem{wang2020channel}
Z.~Wang, L.~Liu, and S.~Cui, ``Channel estimation for intelligent reflecting
  surface assisted multiuser communications: Framework, algorithms, and
  analysis,'' \emph{IEEE Transactions on Wireless Communications}, vol.~19,
  no.~10, pp. 6607--6620, 2020.

\bibitem{he2019cascaded}
Z.-Q. He and X.~Yuan, ``Cascaded channel estimation for large intelligent
  metasurface assisted massive {MIMO},'' \emph{IEEE Wireless Communications
  Letters}, vol.~9, no.~2, pp. 210--214, 2019.

\bibitem{xia2020intelligent}
S.~Xia and Y.~Shi, ``Intelligent reflecting surface for massive device
  connectivity: Joint activity detection and channel estimation,'' in
  \emph{ICASSP 2020-2020 IEEE International Conference on Acoustics, Speech and
  Signal Processing (ICASSP)}.\hskip 1em plus 0.5em minus 0.4em\relax IEEE,
  2020, pp. 5175--5179.

\bibitem{de2020parafac}
G.~T. de~Ara{\'u}jo and A.~L. de~Almeida, ``{PARAFAC}-based channel estimation
  for intelligent reflective surface assisted {MIMO} system,'' in \emph{2020
  IEEE 11th Sensor Array and Multichannel Signal Processing Workshop
  (SAM)}.\hskip 1em plus 0.5em minus 0.4em\relax IEEE, 2020, pp. 1--5.

\bibitem{yuan2022channel}
J.~Yuan, G.~C. Alexandropoulos, E.~Kofidis, T.~L. Jensen, and E.~De~Carvalho,
  ``Channel tracking for ris-enabled multi-user {SIMO} systems in time-varying
  wireless channels,'' in \emph{2022 IEEE International Conference on
  Communications Workshops (ICC Workshops)}.\hskip 1em plus 0.5em minus
  0.4em\relax IEEE, 2022, pp. 145--150.

\bibitem{wei2021channel}
L.~Wei, C.~Huang, G.~C. Alexandropoulos, C.~Yuen, Z.~Zhang, and M.~Debbah,
  ``Channel estimation for {RIS}-empowered multi-user {MISO} wireless
  communications,'' \emph{IEEE Transactions on Communications}, vol.~69, no.~6,
  pp. 4144--4157, 2021.

\bibitem{ris_training}
B.~Zheng, C.~You, and R.~Zhang, ``Intelligent reflecting surface assisted
  multi-user {OFDMA}: Channel estimation and training design,'' \emph{IEEE
  Trans. Wirel. Commun.}, vol.~19, no.~12, pp. 8315--8329, 2020.

\bibitem{ris_channel_estimation_lmmse}
N.~K. Kundu and M.~R. McKay, ``Channel estimation for reconfigurable
  intelligent surface aided {MISO} communications: From {LMMSE} to deep
  learning solutions,'' \emph{IEEE Open J. Commun. Soc.}, vol.~2, pp. 471--487,
  2021.

\bibitem{ris_joint_training_phase}
J.-M. Kang, ``Intelligent reflecting surface: Joint optimal training sequence
  and refection pattern,'' \emph{IEEE Commun. Lett.}, vol.~24, no.~8, pp.
  1784--1788, 2020.

\bibitem{wu2023parametric}
J.~Wu, S.~Kim, and B.~Shim, ``Parametric sparse channel estimation for
  {RIS}-assisted terahertz systems,'' \emph{IEEE Trans. Commun.}, vol.~71,
  no.~9, pp. 5503--5518, 2023.

\bibitem{he2021channel}
J.~He, H.~Wymeersch, and M.~Juntti, ``Channel estimation for {RIS}-aided
  {mmWave MIMO} systems via atomic norm minimization,'' \emph{IEEE Trans.
  Wirel. Commun.}, vol.~20, no.~9, pp. 5786--5797, 2021.

\bibitem{liu2020matrix}
H.~Liu, X.~Yuan, and Y.-J.~A. Zhang, ``Matrix-calibration-based cascaded
  channel estimation for reconfigurable intelligent surface assisted multiuser
  {MIMO},'' \emph{IEEE J. Sel. Areas Commun.}, vol.~38, no.~11, pp. 2621--2636,
  2020.

\bibitem{zhou2022channel}
G.~Zhou, C.~Pan, H.~Ren, P.~Popovski, and A.~L. Swindlehurst, ``Channel
  estimation for {RIS}-aided multiuser millimeter-wave systems,'' \emph{IEEE
  Trans. Signal Process.}, vol.~70, pp. 1478--1492, 2022.

\bibitem{haghshenas2023parametric}
M.~Haghshenas, P.~Ramezani, M.~Magarini, and E.~Bj{\"o}rnson, ``Parametric
  channel estimation with short pilots in {RIS}-assisted near-and far-field
  communications,'' \emph{arXiv preprint arXiv:2308.10668}, 2023.

\bibitem{zhu2023channel}
Y.~Zhu, Y.~Liu, Q.~Wu, C.~You, and Q.~Shi, ``Channel estimation by transmitting
  pilots from reconfigurable intelligent surface,'' \emph{IEEE Transactions on
  Wireless Communications}, 2023.

\bibitem{long2023channel}
W.-X. Long, M.~Moretti, L.~Sanguinetti, and R.~Chen, ``Channel estimation in
  {RIS}-aided communications with interference,'' \emph{IEEE Wireless Commun.
  Lett.}, 2023.

\bibitem{Torres2021}
A.~d.~J. Torres, L.~Sanguinetti, and E.~Björnson, ``Electromagnetic
  interference in {RIS}-aided communications,'' \emph{IEEE Wireless Commun.
  Lett.}, pp. 1--1, 2021.

\bibitem{chandra2022downlink}
G.~S. Chandra, R.~K. Singh, S.~Dhok, P.~K. Sharma, and P.~Kumar, ``Downlink
  {URLLC} system over spatially correlated {RIS} channels and electromagnetic
  interference,'' \emph{IEEE Wireless Commun. Lett.}, vol.~11, no.~9, pp.
  1950--1954, 2022.

\bibitem{vega2022physical}
J.~D. Vega-S{\'a}nchez, G.~Kaddoum, and F.~J. L{\'o}pez-Mart{\'\i}nez,
  ``Physical layer security of {RIS}-assisted communications under
  electromagnetic interference,'' \emph{IEEE Communications Letters}, vol.~26,
  no.~12, pp. 2870--2874, 2022.

\bibitem{hassouna2023reconfigurable}
S.~Hassouna, M.~A. Jamshed, M.~Ur-Rehman, M.~A. Imran, and Q.~H. Abbasi,
  ``Reconfigurable intelligent surfaces aided wireless communications with
  electromagnetic interference,'' in \emph{2023 17th European Conference on
  Antennas and Propagation (EuCAP)}, 2023.

\bibitem{long2023mmse}
W.-X. Long, M.~Moretti, A.~Abrardo, L.~Sanguinetti, and R.~Chen, ``{MMSE}
  design of {RIS}-aided communications,'' \emph{arXiv preprint
  arXiv:2312.07864}, 2023.

\bibitem{khaleel2023electromagnetic}
A.~Khaleel and E.~Basar, ``Electromagnetic interference cancellation for
  {RIS}-assisted communications,'' \emph{IEEE Commun. Lett.}, 2023.

\bibitem{demir2022channel}
{\"O}.~T. Demir, E.~Bj{\"o}rnson, and L.~Sanguinetti, ``Channel modeling and
  channel estimation for holographic massive {MIMO} with planar arrays,''
  \emph{IEEE Wireless Commun. Lett.}, vol.~11, no.~5, pp. 997--1001, 2022.

\bibitem{massivemimobook}
E.~Bj\"{o}rnson, J.~Hoydis, and L.~Sanguinetti, ``Massive {MIMO} networks:
  {Spectral}, energy, and hardware efficiency,'' \emph{Foundations and
  Trends{\textregistered} in Signal Processing}, vol.~11, no. 3-4, pp.
  154--655, 2017.

\bibitem{Bjornson2021b}
E.~{Björnson} and L.~Sanguinetti, ``Rayleigh fading modeling and channel
  hardening for reconfigurable intelligent surfaces,'' \emph{IEEE Wireless
  Commun. Lett.}, vol.~10, no.~4, pp. 830--834, 2021.

\bibitem{Sayeed2002a}
A.~Sayeed, ``Deconstructing multiantenna fading channels,'' \emph{{IEEE} Trans.
  Signal Process.}, vol.~50, no.~10, pp. 2563--2579, 2002.

\bibitem{Demir2021RIS}
{\"O}.~T. Demir and E.~{Björnson}, ``Is channel estimation necessary to select
  phase-shifts for {RIS}-assisted massive {MIMO}?'' \emph{IEEE Trans. Wirel.
  Commun.}, vol.~21, no.~11, pp. 9537--9552, 2022.

\bibitem{Shiu2000a}
D.~Shiu, G.~Foschini, M.~Gans, and J.~Kahn, ``Fading correlation and its effect
  on the capacity of multielement antenna systems,'' \emph{{IEEE} Trans.
  Commun.}, vol.~48, no.~3, pp. 502--513, 2000.

\bibitem{swindlehurst2022channel}
A.~L. Swindlehurst, G.~Zhou, R.~Liu, C.~Pan, and M.~Li, ``Channel estimation
  with reconfigurable intelligent surfaces—a general framework,''
  \emph{Proceedings of the IEEE}, vol. 110, no.~9, pp. 1312--1338, 2022.

\bibitem{Kay1993a}
S.~M. Kay, \emph{Fundamentals of statistical signal processing: Estimation
  theory}.\hskip 1em plus 0.5em minus 0.4em\relax Prentice Hall, 1993.

\bibitem{huang2022semi}
C.~Huang, J.~Xu, W.~Zhang, W.~Xu, and D.~W.~K. Ng, ``Semi-blind channel
  estimation for {RIS}-assisted {MISO} systems using expectation
  maximization,'' \emph{IEEE Transactions on Vehicular Technology}, vol.~71,
  no.~9, pp. 10\,173--10\,178, 2022.

\bibitem{Kotecha2004a}
J.~Kotecha and A.~Sayeed, ``Transmit signal design for optimal estimation of
  correlated {MIMO} channels,'' \emph{{IEEE} Trans. Signal Process.}, vol.~52,
  no.~2, pp. 546--557, 2004.

\bibitem{Liu2007a}
Y.~Liu, T.~Wong, and W.~Hager, ``Training signal design for estimation of
  correlated {MIMO} channels with colored interference,'' \emph{{IEEE} Trans.
  Signal Process.}, vol.~55, no.~4, pp. 1486--1497, 2007.

\bibitem{Bjornson2010a}
E.~Bj{\"{o}}rnson and B.~Ottersten, ``A framework for training-based estimation
  in arbitrarily correlated {Rician} {MIMO} channels with {Rician}
  disturbance,'' \emph{{IEEE} Trans. Signal Process.}, vol.~58, no.~3, pp.
  1807--1820, 2010.

\bibitem{liu2022joint}
Y.~Liu, Q.~Shi, Q.~Wu, J.~Zhao, and M.~Li, ``Joint node activation, beamforming
  and phase-shifting control in {IoT} sensor network assisted by reconfigurable
  intelligent surface,'' \emph{IEEE Transactions on Wireless Communications},
  vol.~21, no.~11, pp. 9325--9340, 2022.

\bibitem{he2024joint}
Z.~He, J.~Xu, H.~Shen, W.~Xu, C.~Yuen, and M.~Di~Renzo, ``Joint training and
  reflection pattern optimization for non-ideal {RIS}-aided multiuser
  systems,'' \emph{IEEE Transactions on Communications}, 2024.

\bibitem{kay_complex_crlb}
V.~Nagesha and S.~Kay, ``{Cramer-Rao} lower bounds for complex parameters,''
  \url{https://www.ele.uri.edu/faculty/kay
  /New\%20web/downloadable\%20files/Nageha\_complex\%20CRLB.pdf}, 2024,
  accessed: 2024-08-06.

\bibitem{fara2021reconfigurable}
R.~Fara, D.-T. Phan-Huy, P.~Ratajczak, A.~Ourir, M.~Di~Renzo, and J.~De~Rosny,
  ``Reconfigurable intelligent surface-assisted ambient backscatter
  communications--experimental assessment,'' in \emph{2021 IEEE international
  conference on communications workshops (ICC Workshops)}.\hskip 1em plus 0.5em
  minus 0.4em\relax IEEE, 2021, pp. 1--7.

\bibitem{Wolff2023Continuous}
A.~Wolff, L.~Franke, S.~Klingel, J.~Krieger, L.~Mueller, R.~Stemler, and
  M.~Rahm, ``Continuous beam steering with a varactor-based reconfigurable
  intelligent surface in the ka-band at 31 {GHz},'' \emph{Journal of Applied
  Physics}, 2023.

\end{thebibliography}

\end{document}